\newif\ifFull
\Fullfalse

\newif\ifri
\rifalse

\newif\ifOurBaseline
\OurBaselinetrue

\newif\ifNewNumbers
\NewNumberstrue

\newif\ifIncludeMin
\IncludeMinfalse

\newif\ifIncludeMax
\IncludeMaxfalse

\newif\ifAppendix
\Appendixfalse

\newif\ifImageNote
\ImageNotefalse    

\newif\ifSuggestedImage
\SuggestedImagetrue   

\newif\ifrids
\ridsfalse

\documentclass[10pt]{article}
\usepackage{fullpage}

\ifFull
\usepackage {url,hyperref}
\fi{}
\usepackage {amsmath}
\usepackage {amssymb}
\usepackage {color}
\usepackage[usenames,dvipsnames,svgnames,table] {xcolor}
\usepackage {wrapfig}
\usepackage {graphicx}
\usepackage {xspace}
\usepackage {floatrow}
\usepackage {subcaption}
\usepackage {float}
\usepackage {tabularx}
\usepackage{booktabs}
\usepackage{numprint}
\usepackage{pdfpages}
\npdecimalsign{.}

\usepackage[normalem]{ulem}
\usepackage{xpatch}
\xpatchcmd{\sout}
  {\bgroup}
  {\bgroup}
  {}{}

\newcommand{\stkout}[1]{\ifmmode\text{\sout{\ensuremath{#1}}}\else\sout{#1}\fi}

\usepackage[final]{listings}

\newcommand {\etal}{{et al.}}

\newcommand {\eg}{{e.g., }}
\ifFull
\newcommand {\fig}{{Figure}}
\else
\newcommand {\fig}{{Fig.}}
\fi

\newcommand {\NP} {$\mathcal {NP}$}

\definecolor {infocolor} {rgb} {0.6,0.6,0.6}

\definecolor {sepia} {rgb} {0.75,0.30,0.15}

\newcommand{\marrow}{\marginpar[\hfill$\longrightarrow$]{$\longleftarrow$}}
\newcommand{\beautifulremark}[3]{\textcolor{blue}{\textsc{#1 #2:}}
\textcolor{red}{\marrow\textsf{#3}}}

\newcommand{\darren}[2][says]{\beautifulremark{Darren}{#1}{#2}}
\newcommand{\imagenote}[1]{{\ifImageNote \color{OliveGreen} Image change: }{\color{RedOrange} #1 \fi}}

\title{Shared Memory Parallel Subgraph Enumeration}

\author{
Raphael Kimmig\thanks{Institute of Theoretical Informatics, Karlsruhe Institute of Technology (KIT), Germany;} \and Henning Meyerhenke\thanks{Institute of Theoretical Informatics, Karlsruhe Institute of Technology (KIT), Germany; {\small \texttt{meyerhenke@kit.edu}}} \and Darren Strash\thanks{Department of Computer Science, Colgate University, USA; {\small \texttt{dstrash@cs.colgate.edu}}}
}

\begin{document}
\maketitle
\begin{abstract}
The subgraph enumeration problem asks us to find all subgraphs of a target graph
that are isomorphic to a given pattern graph.
Determining whether even one such isomorphic subgraph exists is \NP-complete---and
therefore finding all such subgraphs (if they exist) is a time-consuming task.
Subgraph enumeration has applications in many fields, including biochemistry and social networks, and interestingly
the fastest algorithms for solving the problem for biochemical inputs are sequential. 
Since they depend on depth-first tree traversal, an efficient parallelization is far from trivial. Nevertheless,
since important applications produce data sets with increasing difficulty,
parallelism seems beneficial. 

We thus present here a shared-memory parallelization of 
the state-of-the-art subgraph enumeration algorithms RI and RI-DS
(a variant of RI for dense graphs) by Bonnici et al. [BMC Bioinformatics, 2013]. Our strategy uses work stealing and our implementation
demonstrates a significant speedup on real-world biochemical
data---despite a highly irregular data access pattern. 
We also improve RI-DS by pruning the search space better; this
further improves the empirical running times compared to the already highly tuned RI-DS\@.\\

\end{abstract}

\section{Introduction}
\label{sec:Introduction}
%
Graphs are used in a plethora of fields to model relations or interactions between entities.
One frequently occurring (sub)task in graph-based analysis
is the \emph{subgraph isomorphism problem} (SGI).
It requires finding a smaller \emph{pattern graph} $G_p$
in a larger \emph{target graph} $G_t$ or, equivalently, finding an injection
from the nodes of $G_p$ to the nodes of $G_t$ such that the edges of $G_p$ are
preserved.
The SGI decision problem is \NP-complete~\cite{Read1977}. Finding \emph{all} isomorphic subgraphs
of $G_p$ in $G_t$ is commonly referred to as the \emph{subgraph enumeration} (SGE) problem;
this is the problem we deal with in this paper. Note that SGE algorithms
require exponential time in the worst case, as there may be exponentially many matches to enumerate.

Efficient algorithms for SGI exist only for special cases such as planar graphs~\cite{Eppstein1999}, where a linear time algorithm exists for constant query graph size; this method can further be used to \emph{count} the number of occurrences of the query graph in linear time. Consequently, the most successful tools in practice for more general graphs are quite time-consuming when one or all exact subgraphs are sought~\cite{DBLP:journals/pvldb/LaiQLC15}. 
At the same time, the data volumes in common SGI and SGE applications are steadily increasing.
Example fields are life science~\cite{Bonnici2013}, 
complex network analysis~\cite{DBLP:journals/pvldb/LaiQLC15,Erciyes:2014:CNA:2678067}, 
decompilation of computer programs~\cite{Liu2016}, 
and computer vision~\cite{LeBodic2012}. 
A problem related to SGE is \emph{motif discovery}, where the goal is to find all 
frequent subgraphs up to a very small size~\cite{Grochow2007}; note that
in our problem, queries consist of only one subgraph.
Similar to motif search, works dealing with
massive target graphs often focus on very small pattern graphs~\cite{Sun:2012:ESM:2311906.2311907}, often with only up to 10-20
vertices. In contrast, we enumerate pattern graphs with several dozens of vertices and hundreds of edges.

In bioinformatics and related life science fields, graphs are used among other things for analyzing protein-protein
interaction networks and finding chemical similarities~\cite{Dahm2014}.
Such data is often labeled, which further speeds up SGE algorithms by excluding from search those vertex pairs with different labels.
The fastest algorithms for SGE on these graphs use a backtracking approach based
on depth-first search (DFS) of the search space~\cite{Lee2013},
where often pruning rules are employed to reduce the search space. 
The fastest such algorithm is RI by Bonnici \etal~\cite{Bonnici2013}, as evidenced
by a recent study by Carletti \etal~\cite{Carletti2013}.
RI, as well as other state-of-the-art algorithms, are not parallel and thus
there is potential for solving large and/or hard instances faster by parallelization.
Distributed SGE algorithms exist (\eg for MapReduce~\cite{DBLP:journals/pvldb/LaiQLC15}).
However, when the target graph fits into main memory (which is typically the case for the mentioned life science
applications), a shared-memory parallelization is much more promising.
Parallelizing DFS-based backtracking efficiently is not trivial 
though~\cite{Schmidt2009417}.
As in our case, this usually stems from a highly irregular data access pattern, which makes load balancing difficult (in particular when enumerating \emph{all} isomorphic subgraphs).

\paragraph*{Contribution}
Our contribution is twofold. First, we present a shared memory parallelization of subgraph enumeration algorithms 
RI and
RI-DS (the version of RI for dense graphs) using work stealing with private
double-ended queues (see Section~\ref{sec:parallel-ri}).
We conduct a detailed experimental analysis on three data collections, consisting of fifty target graphs and thousands
of pattern graphs from the original RI paper~\cite{Carletti2013}. On
these three data collections we achieve parallel speedups of 5.96, 5.21, and 9.49 with 16 workers on long running instances,
respectively (see Section~\ref{sec:Evaluation}). The maximum speedup for any single instance is 13.40. 
Moreover, on one data collection we manage to reduce the
number of instances not solved within the time limit of 180 seconds by more
than 50\%.
Second, we introduce an improved version of RI-DS that makes better use of
available data by further constraining the search space with minimal overhead
(Section~\ref{sec:content:rids}).
It decreases search space size and variability considerably and somewhat reduces the
average running time on the three data collections.
Our combined improvements give considerable performance gains
over state-of-the-art implementations for longer running instances: 
we achieve speedups of 7.75 (RI), 4.37 (RI-DS) and 13.67 (RI-DS),
respectively, over the original implementations on the three data collections used
in our experiments.

\section{Preliminaries}
\label{sec:prelim}
\subsection{Definitions and Notation}
\label{sub:notation}
\paragraph*{Basics}
A graph $G = (V,E)$ consists of a set of $n$ vertices $V$ and a set of $m$ edges $E
\subseteq V \times V$. Unless stated otherwise, we assume all graphs to be directed. 
We refer to the set of nodes that have an edge starting or ending at $v$ by $N(v) \subseteq V$
or by \textit{neighborhood} of $v$. For directed graphs one can also distinguish between 
incoming and outgoing neighbors. 
In an undirected graph the degree of a node is the number of edges incident to that node.
For the directed case we define the \emph{indegree} of a node $v$ as
$\deg^-(v) = |\{u \mid (u,v) \in E\}|$ 
and the \emph{outdegree} as $\deg^+(v) = |\{w \mid (v, w) \in E\}|$.

\paragraph*{Labels}
Graphs can be annotated with semantic information by adding labels to nodes or
edges. Let $L_V$ and $L_E$ be the sets of possible node and edge labels, respectively.
We define the node label function $lab: V \rightarrow L_V$ that associates each node
with a label. Similarly, we define the edge label function $\beta: E \rightarrow L_E$.
We say two nodes $u,v$ are \emph{equivalent} and write $u \equiv v$ if $lab(u) = lab(v)$.
Two edges $e,f$ are \emph{compatible} if $\beta(e) = \beta(f)$.
We assume strict equality for labels but there may be 
other application-specific equivalence functions. 
%

%
%

\paragraph*{(Sub)Graph isomorphism}
Two graphs $G_1 = (V_1, E_1)$ and $G_2 = (V_2, E_2)$ are considered
\emph{isomorphic}, denoted by $G_1 \cong G_2$, if there is a bijective function
$f$ mapping all nodes of $G_1$ to nodes of $G_2$ such that the edges of the
graph are preserved and compatible, and nodes in $G_1$ are mapped onto equivalent
nodes in $G_2$.  More formally, $G_1 \cong G_2$ if and only if a bijection $f:
V_1 \rightarrow V_2$ exists that fulfills the following properties.
\label{eq:gi}
\begin{equation} \label{eq:gi_preserve_structure}
\forall u,v \in V_1: (u,v) \in E_1 \iff (f(u), f(v)) \in E_2
\end{equation}
\vspace{-3.5ex}
\begin{equation} \label{eq:gi_preserve_node_labels}
    \forall v \in V_1: v \equiv f(v)
\end{equation}
\vspace{-3.5ex}
\begin{equation} \label{eq:gi_preserve_edge_labels}
    \forall (u,v) \in E_1: (u,v) \equiv (f(u), f(v))
\end{equation}
In \emph{subgraph enumeration} we must list all (in our case non-induced) subgraphs in a target graph $G_t$ that 
are isomorphic to a pattern graph $G_p$. 
Clearly, subgraph enumeration is not 
easier than subgraph isomorphism. In practice one can even expect a significant slowdown,
even when working in parallel. After all, we cannot stop after a potentially
early first hit, but have to explore the search space exhaustively.

\subsection{Related Work}
\label{sec:RelatedWork}
\subsubsection{Overview}
According to Carletti \etal~\cite{Carletti2015}, most SGI and SGE algorithms can be categorized as follows.


\paragraph{State space based} 
A common way to model the SGI/SGE search space is to use a \textit{state space representation}
(SSR)~\cite{Carletti2013}, modeling the search space as a tree.
Finding a subgraph isomorphism can then be viewed as the problem
of finding a mapping $M: V_p \rightarrow V_t$ that maps nodes of the pattern
graph
onto nodes of the target graph. If the mapping is injective and does not
violate the isomorphism constraints
\eqref{eq:gi_preserve_structure} to~\eqref{eq:gi_preserve_edge_labels}, it yields a subgraph $G_s \subseteq
G_t$ isomorphic to $G_p$. 

    
Every node in the tree represents a state of the (partial) mapping; the root represents the empty mapping.
A branch taken represents extending the partial mapping by a certain pair of nodes. To find a 
subgraph isomorphism, one thus needs to find a path in the state space tree of length $|V(G_p)|$
beginning at the root, such that the equivalent mapping does not violate the
graph isomorphism constraints. The key to doing that efficiently is ignoring
parts of the state space tree early which cannot be extended to a valid solution. This
is especially critical when enumerating \emph{all} isomorphic subgraphs, since otherwise states
at depth $|V(G_p)|$ in the state space tree would need to be visited.
%
%
Typically, state space based approaches use depth first search (DFS) in
order to search the state space tree and make use of pruning rules in order to 
remove parts of the search space that do not contain valid
solutions~\cite{Cordella2004,Lee2013}.
In an independent experimental comparison, Carletti \etal~\cite{Carletti2013} concluded that ``RI seems
to be currently the best algorithm for subgraph isomorphism'' for sparse 
biochemical graphs. Other popular algorithms of this category besides RI~\cite{Bonnici2013} are VF2~\cite{Cordella2004} and VF2 Plus~\cite{Carletti2015} (VF2 was also part of Carletti \etal's comparison).

One important distinction between the different state space exploration
approaches is the \emph{order} in which the nodes of the pattern graph are processed.
Algorithms like VF2 use a \textit{dynamic variable ordering}: they
decide at every state, based on the nodes already mapped onto the target graph,
which node of the pattern graph to examine next. This allows them a greater
freedom in eliminating unfruitful branches of the search space but comes with
the additional cost of running whatever logic is used to pick the next pattern
graph node in every step of the search process.
Algorithms like RI and VF2 Plus use a static ordering fixed before the search.
This reduces the amount of work done during the search process but
means that at any step the selection of the next variable may not be optimal
regarding search space size~\cite{Almasri2015}.
Since we base our parallelization on RI, a more detailed explanation of the sequential algorithm is worthwhile.
As already mentioned, RI uses state space exploration with a search strategy that is static and depends only on the pattern graph. The general
idea is to order the nodes of the pattern graph in a way that ensures the next
node visited is always the one being most constrained by already matched nodes,
while introducing additional constraints as early as possible. During the
search no expensive pruning or inference rules are used, trading faster
comparisons for a larger search space. The pattern graph nodes are ordered by
starting with highly connected nodes and then greedily adding nodes that are
connected with already selected nodes. For the actual search process, a set of
increasingly expensive rules are checked for each candidate extension.

There is also a version of RI called RI-DS~\cite{Bonnici2013} that computes
an initial list of possible target nodes for all nodes in $G_p$ and is better on medium to
large dense graphs~\cite{Bonnici2013,Carletti2013}.
RI and RI-DS participated in the ICPR2014 contest on
graph matching algorithms for pattern search in biological databases,
``outperforming all other methods in terms of running time and memory
consumption''~\cite{Bonnici2014}.

\paragraph{Index based} Index-based approaches employ an indexing data structure to store features (such as distance in the graph) that can then queried to quickly find a valid extension of a partial mapping. This reduces the time to find an isomorphic subgraph if one exists (called the \textit{matching time}). Index-based approaches spend more time up front \textit{preprocessing} the target graph, and are therefore typically used for larger target graphs and very small
pattern graphs. Examples include QuickSI~\cite{Shang2008}, GraphQL~\cite{He2010},
GADDI~\cite{Lee2013} and STwig~\cite{Sun:2012:ESM:2311906.2311907}.

\paragraph{Constraint propagation based} 
Modeling the subgraph isomorphism problem as a constraint satisfaction problem (CSP)
means that each variable/node in $G_p$ has a domain (a set of
candidate nodes in $G_t$) and a set of constraints that ensure edge
preservation and injectivity.
Tools in this category focus on reducing the search space at the cost of spending more time to
achieve that reduction. The search space reduction is dependent on both the pattern
and target graph~\cite{Bonnici2013}.
LAD~\cite{Solnon2010} is a CSP approach in which initial domains can be based solely on node
degrees, but can also incorporate label compatibility. 
In addition to the constraints above, LAD uses a constraint 
which results in the removal of values from the domain if
using them would imply remaining nodes cannot all be mapped to different nodes.
Also, LAD propagates constraints after each assignment to further reduce domains of
remaining variables.
These groups are not disjoint; LAD for example uses a state space
representation with depth first search, but it reduces the search space
by using constraint propagation~\cite{Carletti2013}.

\subsubsection{Parallelism}
Parallel backtracking has been considered in recent years both in a generic manner (\eg \cite{AbuKhzam201565})
and for specific combinatorial search problems such as maximal clique enumeration~\cite{Schmidt2009417}.
To the best of our knowledge, there are no parallel versions of the newer state-of-the-art algorithms for subgraph enumeration or isomorphism
like RI and VF2 Plus.
Some work has been done in order to accelerate the search process using GPUs,
with an algorithm named GPUSI, but here the focus is on tiny pattern graphs~\cite{Yang2015}.
The work by Shahrivari and Jalili~\cite{DBLP:journals/sp/ShahrivariJ15} is also for shared memory, but targets motif search. Thus, there are a few critical differences: (i) they search for all subgraphs of a certain size $k$, (ii) this size $k$ is tiny, and (iii) their parallelization approach is simpler due to the problem structure.
There is also a CSP based parallel approach for SGI that beats VF2 and LAD on some
graphs, but it was not compared to RI nor RI-DS~\cite{McCreesh2015}. 
Of the backtracking approaches there is one parallelization of
VF2~\cite{Blankstein2010} using Cilk++. Its authors note that the
amount of state copied to enable work stealing results in a lot of overhead.
To remedy this, our parallelization copies partial solutions only for stolen tasks, 
not those that remain private.

RI's static node ordering, and focus on speed of
exploration, makes it a good candidate for parallelization. The static
order and lack of complex state should allow for reasonable overhead when distributing
work among workers.

\ifri
\section{RI}
\label{sec:ri}
 We begin by discussing several critical features of RI, which we use in our parallelization.

RI performs fast backtracking search by choosing vertices for inclusion in the solution by using a static \emph{constraint-first variable ordering} of pattern graph nodes, and by employing efficient rules for pruning the search space.

In particular, RI repeatedly expands a mapping $M: G_p \rightarrow G_t$ that reflects the current (partial) solution. 
And the vertex selection order ensures the most \emph{constrained} nodes are chosen first, and that additional constraints are introduced as soon as possible---which significantly reduce the search space.

We denote a (partial) ordering by $\mu = [\mu_0 \ldots \mu_{|E(G_p)|} ]$ where $\mu_i \in G_p$ and $\mu_i \ne \mu_j$ if $i \ne j$. 
The nodes of $G_p$ will be inserted into $M$ in this order.

The search begins with an empty mapping $M$. The first pattern node in $\mu$, $\mu_0$,
is checked against a node $v_t \in G_t$. This check is performed using a number
of pruning rules (see next paragraph) to ascertain whether 
expanding $M$ to include $\mu_0 \rightarrow v_t$ would violate any 
of the subgraph isomorphism constraints. If that is not the case $M$ is
expanded accordingly and the next node to be processed is $\mu_1$.

%
%

\ifFull
\begin{figure}[h]
\floatsetup{capposition = below, floatrowsep =qquad,}
\ffigbox{%
\begin{subfloatrow}
    \ffigbox[0.5\textwidth]{\caption{Nodes 1, 2 and 3 have one neighbor in $\mu$ each so the one with the
        most nodes in $\mu$ reachable via non-$\mu$ nodes is selected (node 1).}}{%
        \includegraphics[width=0.4\textwidth]{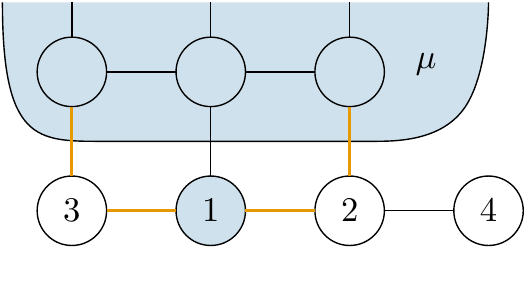}%
  }
    \ffigbox[0.5\textwidth]{\caption{Nodes 2 and 3 both have two neighbors in
        $\mu$ and zero neighbors that themselves are neighbors of $\mu$, so the one
    with higher degree is selected (node 2).}}{%
    \includegraphics[width=0.4\textwidth]{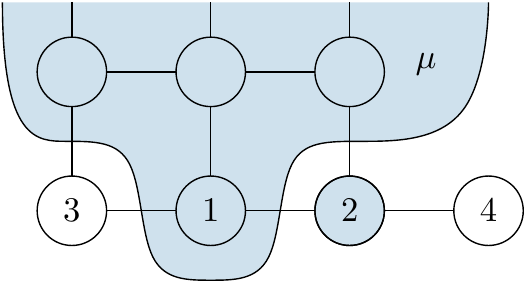}%
  }
\end{subfloatrow}
\begin{subfloatrow}
    \ffigbox[0.5\textwidth]{\caption{Node 3 has two neighbors in $\mu$ while node 4 has only one, so node 3 is selected.}}{%
        \includegraphics[width=0.4\textwidth]{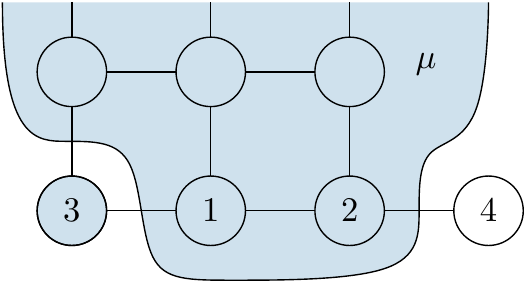}%
  }
    \ffigbox[0.5\textwidth]{\caption{Node 4 is selected.}}{%
        \includegraphics[width=0.4\textwidth]{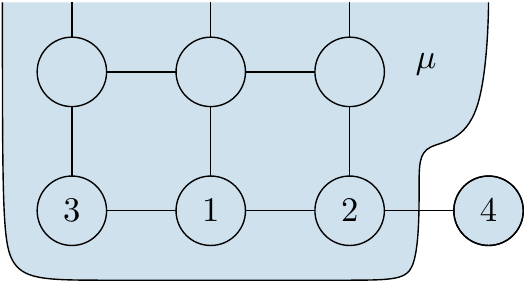}%
  }
\end{subfloatrow}
}{%
\caption{The constraint-first node ordering of RI, nodes in $\mu$
are already ordered.}
\label{fig:node_ordering}
}
\end{figure}
\fi

We can illustrate how search proceeds using the vertex ordering by growing a mapping $M: G_p \rightarrow G_t$.



RI begins by selecting the node with the highest degree and adding it to $\mu$
resulting in $\mu = [\mu_0]$.
Afterwards, it greedily adds nodes until all nodes of $G_p$ are in $\mu$.
Let $\mu$ be a partial variable ordering and $v_p$ be a pattern node.
We denote by $w_m(\mu, v_p)$ the number of neighbors of $v_p$ in $\mu$ 
and by $w_n(\mu, v_p)$ the number nodes in $\mu$ reachable via nodes not in $\mu$.
Formally we define $w_m(\mu, v_p) = |N(v_p) \cup \mu|$ and 
$w_n(\mu, v_p) = |\{w \in \mu \mid \exists x \notin \mu: \{v_p, w\} \subseteq N(x)\}|$.
For the purpose of the ordering we do not care about the directionality of
edges.

In each iteration the algorithm picks the unprocessed node with the most neighbors in
$\mu$, that is the one with the highest value $w_m$. This is the \emph{fail-first principle} in action: picking the most constrained node first. If multiple nodes have the
same $w_m$, ties are broken first by the highest number of nodes in $\mu$ that are reachable
via nodes not yet in $\mu$, then by highest degree.\footnote{This differs from the description in~\cite{Bonnici2013}, but matches the implementation of RI 3.6.} 
 That is the one with the highest $w_n$ is picked---maximizing the number of further constraints applied.


During the construction process the algorithm also keeps track of, for each
node, the first node in the ordering from which it can be reached. The resulting parent
mapping $P$ is used in the depth first search to select candidate nodes for a
given pattern node.


\ifFull
\begin{figure}
    \begin{lstlisting}[gobble=4, language=Python]
    def constraint_first_ordering(graph):
        """
        Input:
            graph - The pattern graph to be ordered.
        
        Returns:
            ordering - Ordered nodes of the pattern graphs.
            parents  - Mapping of each node to the first node in the ordering
                       from which it can be reached or None.
        """
        remaining = graph.nodes

        ordering = []
        ordering.push(remaining.sort_by(degree).pop_first())
        while not remaining.is_empty():
            remaining = remaining.sort_by(
                number_of_core_neighbors,
                number_of_core_nodes_reachable_through_non_core_nodes,
                degree
            )
            ordering.push(remaining.pop_first())

        parents = {}
        for node in ordering:
            parents[node] = ordering.filter(has_edge_to(node)).first()

        return ordering, parents
    \end{lstlisting}
    \caption{Algorithm for the constraint-first node ordering.}\label{fig:node_ordering_code}
\end{figure}
\fi

At any point in the search process, when processing $\mu_i$, the candidate node
$v_t \in G_t$ is one of the neighbors of $M(P[\mu_i])$ in $G_t$. 
By definition, $P(\mu_i)$ is the first node in $\mu$ from which $\mu_i$ can be reached in $G_p$.
This means that if constraint~\eqref{eq:gi_preserve_structure} holds $M(\mu_i)$
will have to be reachable from $M(P(\mu_i))$ in $G_t$. 
For the first node, and nodes not connected to already mapped nodes in $M$, all
nodes of $G_t$ are candidates.

If none of the pruning rules are violated the mapping is expanded so that
$M(\mu_i) = v_t$. Whenever the mapping reaches the size of $V(G_p)$ it induces
a valid subgraph and a match is reported.


\subsubsection{Pruning rules}
The rules for pruning infeasible paths are kept very simple in order to
minimize the time spent exploring a single state.
There are four rules and they are executed in order, from cheapest to most
costly and evaluation stops as soon as one fails. For a mapping $M$, a pattern
node $\mu_i$ and a target node $v_t$ they ensure the following.
\begin{enumerate}
    \item $v_t$ is not used in the mapping $M$.
    \item  $\mu_i \equiv v_t$ verifying constraint~\eqref{eq:gi_preserve_node_labels}.
    \item Indegree and outdegree of $\mu_i$ are not bigger than indegree and
    outdegree of $v_t$.
    \item Constraint~\eqref{eq:gi_preserve_structure}, that is the structure of
        $G_p$, is preserved and the edge labels are compatible as per
        constraint~\eqref{eq:gi_preserve_edge_labels}.
\end{enumerate}
\fi


\section{Parallelizing RI}
\label{sec:parallel-ri}

Perhaps the biggest challenge for fine-grained parallelism is to keep workers from \emph{starving}. For subgraph isomorphism and enumeration, starvation is even more of a concern, as the search space is highly irregular~\cite{Cong2008,Fatta2007}---few long overlapping search paths exist that lead to solutions. However, there is another challenge: assuming we can find enough tasks for all workers, we need to efficiently communicate tasks, which include a potentially large mapping $M$.   We address both of these challenges by combining the \emph{work stealing} strategy of Acar~\etal~\cite{Acar2013} and \emph{task coalescing}. 

\subsection{Task Representation}
The search phase of RI iteratively expands a mapping $M$ of nodes from $G_p$ onto nodes from $G_t$ by choosing vertices according to a static ordering $\mu = [\mu_1,\ldots,\mu_{V(G_P)}]$ of the pattern nodes. A natural way to represent a task then would be as a tuple $(M, \mu_i, v_t)$ with the current partial mapping $M$, and a check to perform: can we map pattern node $\mu_i$ to target node $v_t$? If the new mapping is consistent, then the task spawns new tasks $(M^\prime, \mu_{i+1}, v_c)$ to check if the next pattern node $\mu_{i+1}$ maps to each remaining candidate target node $v_c$.
%
While this seems like the ideal representation, it has serious drawbacks.
First, the mapping $M$ is potentially large, as it must include every mapped node of $G_p$. Second, when executing a task, a worker explores only one state in the search space. Thus, this task representation comes with large overhead: copying $M$ for each new task is too time consuming.
For that reason, we design our tasks with several built-in optimizations.
First, we do not explicitly store a partial mapping with a task, we instead communicate partial mappings \emph{as needed}.
Therefore, we effectively represent a task by the node pair $(\mu_i, v_t)$.
Second, when expanding a mapping $M$ to include pattern node $\mu_i$ and target node $v_t$,
we first check the consistency of each new task before spawning it. This reduces
the risk of a worker stealing a dead-end task.



\subsection{Work stealing with private deques}
\begin{figure}[]
\centering\includegraphics[]{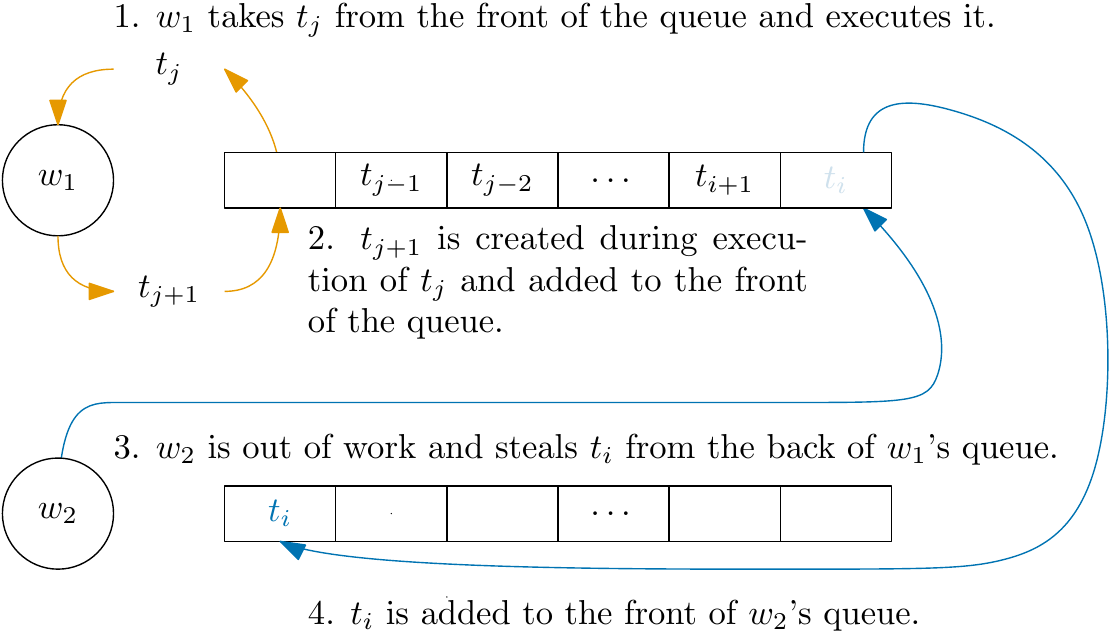}
\caption{Workers $w_1$ and $w_2$ and their deques. $w_1$ is executing a task and $w_2$ is stealing from $w_1$.}
\label{fig:work_stealing}
\end{figure}

Our solution is to use the work stealing strategy of Acar~\etal~\cite{Acar2013}, which gives each worker $i$ a private double-ended queue (deque) $q_i$. Worker $i$ adds and removes from the front of its private deque $q_i$, and workers request to steal from the back of other workers' deques. The private deque serves two purposes for our problem:
(i) Tasks at the front of worker $i$'s deque are added and removed in depth-first search order, and therefore worker $i$ always has a correct partial mapping $M_i$. Thus, for private tasks, a partial mapping is never copied. 
(ii) Tasks at the back of a worker's deque are closer to the root of the search space tree than those near the front. Therefore, when stealing, a worker receives a task with larger part of the tree below them than nodes closer to the leaves. Thus, we expect stolen tasks to be relatively long-running, reducing the number of steals overall.
%
An example of workers stealing and executing tasks can be seen in
\fig~\ref{fig:work_stealing}.

We quickly point out that an alternative solution would be to use \emph{lock-free} data structures, which ensure efficient access to tasks for all workers with low
overhead~\cite{Chase2005, Dijk2014, Faxen2010}. 
However, they have two drawbacks: first they are notoriously difficult to implement~\cite{Acar2013}, and second, they still require copying a partial map with every task.

\ifFull
The straightforward methods to address these drawbacks are not ideal. 
Synchronizing access to the mapping is not a viable option, as the mapping is used in every iteration of the search, and would add high
synchronization overhead~\cite{Hendler2002}. 
Another option is to remove the mapping $M$ from the task, which would
reduce the task to tuple of nodes $(\mu_i, v_t)$. To
check if $\mu_i$ may be mapped to $v_t$, a worker would
still need access to $M$. One way to solve this is by creating
copies of $M$ at each depth of the search and copying those again when
transferring tasks between workers. This is the approach used by Blankstein and 
Goldstein in their VF2 parallelization~\cite{Blankstein2010}.
A further option is to increase the task size so that a worker explores more states (e.g., one thousand) to combat excessive copying.
This would however force us to have less granularity, which is less desirable given the
irregular search space~\cite{Cong2008,Fatta2007}.
\fi

\ifFull
While there are implementations in frameworks like OpenMP that allow easy
parallelization of task-based problems, they can exhibit poor performance with
fine grained tasks~\cite{Olivier2009} and their performance characteristics can
be hard to understand~\cite{Podobas2010}. 
In order to give us full control over all parameters we choose to implement work
stealing ourselves. Many work stealing algorithms rely on lock-free data
structures to ensure efficient access to tasks for all workers with low
overhead~\cite{Chase2005, Dijk2014, Faxen2010}. These data structures are
however notoriously difficult to implement~\cite{Acar2013}.
\fi



Load balancing could either be sender or receiver initiated and is performed by 
all workers explicitly calling communication methods in their work loops.
We implement a receiver-initiated private deque work stealing
since its performance is comparable to classic
work stealing. Aside from a private deque for each worker, this method
requires three shared data structures:

\begin{enumerate}
\item \texttt{work\_available}---an array of Boolean values, one for each worker,
indicating if that worker currently has tasks in its queue.
\item \texttt{requests}---an array of worker ids, used by workers to request a task
from another worker.
\item \texttt{transfers}---an array of tasks, used
to transfer tasks between workers.
\end{enumerate}

\begin{figure}
    \begin{lstlisting}[gobble=8, language=Python]
        while not terminated:
            if q.is_empty():
                acquire_task(worker)
            task = q.pop()
            work_available[worker] = not q.is_empty()
            process_task_requests(worker) 
            execute(task)
    \end{lstlisting}
    \caption{The work loop for work stealing with private deques.}\label{fig:work_stealing_loop}
\end{figure}

The main loop of a worker consists of taking a task from the deque, updating
the entry in \texttt{work\_available}. It then checks for a work request in
\texttt{requests}, answering that via \texttt{transfers} from the back of its
queue if possible and concludes by executing the task it took at the start of
the loop. Once it runs out of tasks, it repeatedly requests work from a random
worker until it receives a task or is terminated. 
The main loop can be seen in
\fig~\ref{fig:work_stealing_loop}.
\ifFull
For synchronization the array
\texttt{requests} contains C++11's \texttt{std::atomic<int>}s to store the
worker ids and \texttt{std::atomic\_compare\_exchange\_weak} is used to ensure
only one request is placed for one worker at any time. No other synchronization
efforts are required.
\fi
%
Function \texttt{process\_task\_requests} transfers partial mappings between workers.  This design keeps tasks small and reduces overhead when creating tasks. Except for the \texttt{requests}\footnote{For \texttt{requests}, we use C++11's \texttt{std::atomic<int>} for each worker id
and \texttt{std::atomic\_compare\_exchange\_weak} to ensure
only one request is placed for one worker at any time.}, all data structures are completely unsynchronized.

\subsection{Initial work distribution}
We initially create tasks corresponding to states
directly below the root of the search space tree. Each task 
maps the node $\mu_1$ of the node ordering onto one node of the target graph.
At the beginning of the search process, each worker creates an equal number
of those tasks and places them in its private deque.

\subsection{Task coalescing}
One further reason to use work stealing with private deques, is the ease with which we can perform \emph{task coalescing}~\cite{Acar2013}---that is, grouping together tasks into \emph{task groups} that can each be processed as a single unit of work. In general, grouping together tasks that have short execution time can reduce the number of steals---and the number of times we must copy a worker's partial mapping.
This design makes it easy to experiment with the size of task groups to strike a balance between overhead and granularity.


\subsection{Termination detection}
Finally, we implement a termination detection algorithm, since there is no central scheduler, and we do not know the number of tasks in advance.
To detect when all workers terminate, we implement a variant of Dijkstra's popular termination algorithm~\cite{Dijkstra1983} described by Schnitger~\cite{panda-schnitger-2009}: workers are arranged in a ring and when a worker becomes idle, it passes a token to the next worker. Initially the token is colored white; if a worker is busy, it colors the token black. If the token makes it around the ring and is still white, then all workers terminate.
%
The algorithm has termination delay proportional to the number of workers. As we use no more than 16 workers for our shared memory implementation, this simple approach is sufficient. More efficient algorithms exist for systems with many more workers or more complex topologies~\cite{Mahapatra2007}.

\section{Speeding up RI-DS}\label{sec:content:rids} 
RI-DS is a variation of RI that is faster on dense graphs~\cite{Bonnici2013} and 
differs from RI by precomputing sets of compatible nodes for each pattern node. These sets are incorporated into the initial node ordering and consistency checks.
Before computing the node ordering, RI-DS first computes for each pattern
node $v_p$ a set of compatible target nodes $D(v_p) \subseteq G_t$ called
the \emph{domain} of $v_p$. This process is called \emph{domain assignment}.

\subsection{Domains in RI-DS}
Initially, the domain of $v_p$, denoted by $D(v_p)$, is set to all pattern
nodes with compatible degrees and equivalent labels. That is,
all nodes with in- and outdegree at least that of $v_p$'s, and with labels that match $v_p$'s.

We then remove each $v_t$ from $D(v_p)$ whose neighborhood is
not consistent with the neighborhood of $v_p$.
For example, let $v_t \in D(V_p)$ and suppose we have an edge
$(v_p, w_p) \in E(G_p)$; then we want at least one $w_t \in D(w_p)$
such that $(v_p, w_p) \cong (v_t, w_t) \in E(G_t)$. If no such edge exists, then
mapping $v_p$ onto $v_t$ cannot lead to a solution and we can remove
$v_t$ from $D(v_p)$.  If any domain becomes empty, then there are
no isomorphic subgraphs to enumerate. This step is based on arc-consistency (AC) from constraint
programming~\cite{Bonnici2014, Solnon2010}. 



Domains are used in both the preprocessing and search phases. First, when computing the node ordering $\mu$, all pattern nodes with domain size one (called \emph{singleton} domains) are placed at the beginning of the ordering. Further, when initializing the search, RI-DS uses domains as candidates for the root node of the search space (unlike RI, which considers $V(G_t)$). Lastly, during the search, to determine if we can map $\mu_i\in V(G_p)$ onto $v_t\in V(G_t)$, we first check if $v_t\in D(\mu_i)$.

\subsection{RI-DS with improved tie-breaking and forward checking}
\label{sec:content:rids-improved}
We integrate two improvements into the preprocessing phase of RI-DS: 
we use domain size to break ties in the node ordering $\mu$, and
we further reduce domain sizes with \emph{forward checking}.

\subsubsection{Node ordering}
Bonnici and Giugno~\cite{Bonnici2014} provide an extensive comparison of node ordering strategies. They show that, 
while other domain-based orderings occasionally reduce the search space further than RI-DS, RI-DS consistently outperforms other methods.
Our goal is to strengthen the node ordering of RI-DS by further using
domains, without introducing unacceptable overhead.
RI's node ordering (on which RI-DS is based) is
constructed by greedily selecting pattern nodes according to the number of
neighbors in the partial ordering (denoted by $w_m$), the number of nodes in the ordering
reachable via nodes not in the ordering (denoted $w_n$), and the degree.
We propose to further break ties when two nodes have the same degree,
in favor of the node with the smaller domain. That is, when deciding
how to order two nodes with
identical values $w_m$ and $w_n$ and identical
degrees, we select the one with the smaller domain to appear first in the node ordering.
This is a continuation of the constraint-first principle: by preferring
the node with the smaller domain, we pick the node which is most constrained first.



\subsubsection{Forward checking}
In forward checking, a concept in constraint programming, once we assign a value
to a variable (in our case, mapping a pattern node onto a target node), we place
additional restrictions on the remaining unassigned variables.
After assigning a variable, we can remove from the domains of all unassigned
variables those values that will violate a constraint because of this
assignment~\cite{Solnon2010}.

\ifFull
\begin{figure}[]
   \centering
   \caption{Example of forward checking in action. In every round singleton
       domains are removed from other domains until no more changes occur.}\label{fig:forward_checking}
   \begin{tabularx}{\textwidth}{X X X X X}
      \toprule
      Pattern node & \multicolumn{4}{c}{Domains} \\
      \cmidrule(lr){2-5}
       & initial & 1st round & 2nd round & result \\
      \midrule

      $n_0$ \newline
      $n_1$ \newline
      $n_2$ \newline
      $n_3$ \newline
      $n_4$ 

      &

      $\{\uline{0}\}$ \newline
      $\{\uline{4}\}$ \newline
      $\{\uline{0}, 3, 7\}$ \newline
      $\{2, 8, 9\}$ \newline
      $\{\uline{0}, \uline{4}, 7\}$ 

      & 

      $\{0\}$ \newline
      $\{4\}$ \newline
      $\{\stkout{0}, 3, \uline{7}\}$ \newline
      $\{2, 8, 9\}$ \newline
      $\{\stkout{0}, \stkout{4}, \uline{7}\}$ 

      & 

      $\{0\}$ \newline
      $\{4\}$ \newline
      $\{\uline{3}, \stkout{7}\}$ \newline
      $\{2, 8, 9\}$ \newline
      $\{7\}$ 

      & 

      $\{0\}$ \newline
      $\{4\}$ \newline
      $\{3\}$ \newline
      $\{2, 8, 9\}$ \newline
      $\{7\}$ 

      \\

      \bottomrule
   \end{tabularx}
\end{figure}
\fi

In RI-DS, assignments take place only during the search phase---not while computing domains. Yet, we observe that pattern nodes with
singleton domains can only be assigned to a single target node. We 
therefore perform forward checking for all pattern nodes with singleton domains, as  each pattern node will be assigned to its target node in the future. 

The constraint we verify is injectivity. For each pattern node with a
singleton domain, we remove the target node in that domain from the domains of 
all other pattern nodes. We repeat this procedure for any newly introduced 
singleton domains.
In RI, domains are implemented as bitmasks, which we use
to quickly remove singleton domains' contents 
from all other domains.

\ifFull
\begin{figure}[]
\centering\includegraphics{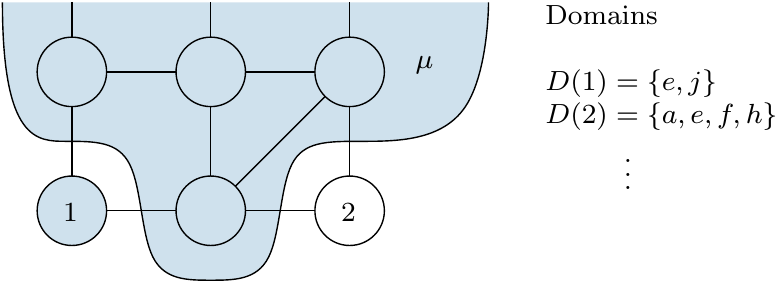}
\caption{The improved node ordering of RI-DS-SI-FC, nodes in $\mu$
are already ordered. The nodes 1 and 2 have equal $w_n$, $w_m$ and degrees.
Node 1 does have a smaller domain and is thus picked first.}
\label{fig:node_ordering_improved}
\end{figure}
\fi

\ifFull
\subsection{Other approaches considered}
Aside from RI and RI-DS as well as our improved version we also 
did some preliminary experiments with several other techniques
hoping that any of them might show good behavior when parallelized.
Among the things we tried where domain size and degree based node orderings
as well as a parallel implementation of VF2 Plus.

Unfortunately none of the orderings showed any real promise and including them
all would have led to a pointless duplication of Bonnici and Giugno's excellent
paper on variable ordering~\cite{Bonnici2014}.

Our implementation of VF2 Plus did not show promising results either. 
We implemented VF2 Plus based on the publicly available version of VF2 and the
paper in which Carletti~\etal~\cite{Carletti2015} introduced VF2 Plus.
On longer running instances (which we focus on) the performance was 
much worse than both RI and RI-DS\@. This is in no way indicative of the
potential performance of VF2 Plus and may be solely due to deficiencies in our
implementation. This may warrant further investigation, but we focus here on
parallelizing RI and RI-DS.
\fi

\section{Experimental Evaluation}
\label{sec:Evaluation}
We now give an in-depth experimental evaluation of our SGE algorithms.
Our experimental evaluation is split into two parts. In the first one we evaluate the
effectiveness of our parallelization. We do so by comparing our parallel RI against
our sequential RI implementation and against the original RI implementation (RI version 3.6). 
In the second part we compare RI-DS (version 3.51) against our new variants with domain size ordering (RI-DS-SI) and with domain size ordering and forward checking (RI-DS-SI-FC).

\ifFull
\begin{table*}[]
   \centering
   \caption{Graph data collections.}\label{tab:datasets}
   \resizebox{\textwidth}{!}{%
   \begin{tabular}{r r r r r r r r r r r r}
      \toprule
      collection& \multicolumn{3}{c}{node count} & \multicolumn{3}{c}{edge count}
      & \multicolumn{2}{c}{degree} & \multicolumn{3}{c}{label count} \\
      \cmidrule(lr){2-4}
      \cmidrule(lr){5-7}
      \cmidrule(lr){8-9}
      \cmidrule(lr){10-12}
      & $\min/\max$ & $\varnothing$ & $\sigma$ 
      & $\min/\max$ & $\varnothing$ & $\sigma$ 
      & $\varnothing$ & $\sigma$
      & $\min/\max$ & $\varnothing$ & $\sigma$ 
      \\
      PPIS32  & \numprint{5720}/\numprint{12575} & \numprint{7827.10} &
      \numprint{2120.15} & \numprint{51464}/\numprint{332458} &
      \numprint{107134.80} & \numprint{82730.88} & 27.38 & 60.88 & 32/32 & 32.00 & 0.00\\
\midrule
GRAEMLIN32  & \numprint{1081}/\numprint{6726} & \numprint{3167.60} &
\numprint{1568.66} & \numprint{12960}/\numprint{230467} & \numprint{87758.70} &
\numprint{75939.11} & 55.41 & 88.74 & 32/32 & 32.00 & 0.00\\
\midrule
PDBSv1 & 240/\numprint{33067} & \numprint{5663.60} & \numprint{6954.82} &
480/\numprint{61546} & \numprint{8661.27} & \numprint{12365.70} & 3.06 & 2.67 & 5/8 & 5.90 & 1.04\\
      \bottomrule
   \end{tabular}}
\end{table*}
\else
\begin{table}[]
\small
   \centering
   \caption{Graph data collections.}\label{tab:datasets}
   \begin{tabular}{@{\hskip -0.5pt}r r r r r@{\hskip -0.5pt}}
      \toprule
      Collection& $|V(G_T)|$ & $|E(G_T)|$ &
      \multicolumn{2}{c}{degree} \\
      & $\min/\max$ &
      $\min/\max$ &
      $\mu$ & $\sigma$ \\
\midrule
      PPIS32  &
      \numprint{5720}/\numprint{12575} &
      \numprint{51464}/\numprint{332458} &
      27.38 & 60.88\\
\midrule
      GRAEMLIN32  &
      \numprint{1081}/\numprint{6726} &
      \numprint{12960}/\numprint{230467} &
      55.41 & 88.74 \\
\midrule
      PDBSv1 &
      240/\numprint{33067} &
      480/\numprint{61546} &
      3.06 & 2.67 \\
      \bottomrule
   \end{tabular}
\end{table}
\fi

\subsection{Data collections}

We select a subset of the six biochemical data collections tested by Bonnici~\etal~\cite{Bonnici2013} for the original experiments with RI. We focus on data collections with large graphs that contain hard, long-running instances, as we do not expect easy instances to benefit from parallelism. 
We select three data collections: PDBSv1, on which RI is more efficient than RI-DS, and PPIS32 and GRAEMLIN32 on which RI-DS is more efficient. 
Properties of these data collections are listed in Table~\ref{tab:datasets} and described next.




\paragraph*{PPI}
The PPI data collection consists of large, dense protein-protein interaction networks, which
have either 32, 64, 28, 256, 512, \numprint{1024}, \numprint{2048}, or unique
labels; For each label count there is a version with a uniform and one with a
normal (Gaussian) distribution.
We run our experiments on the variant with 32 normally distributed labels (which we call PPIS32) since it contains the highest number of unsolved instances.
PPIS32 has ten target graphs;
there are also 420 pattern graphs with 4, 8, 16, 32, 64, 128, and 256 edges, 
which are are classified as either being dense, semi-dense, or sparse~\cite{Bonnici2013}.

\paragraph*{PDBSv1}
The PDBSv1 data collection consists of large, sparse graphs with data from RNA, DNA, and
proteins. 
The 30 target graphs have between 240 and \numprint{330067} nodes. 
There are \numprint{1760} pattern graphs have 4, 8, 16, 32, 64 or 128 edges.
Note that RI could not solve about half of the 128-edge pattern graphs 
in the 3-minute time limit of the original experiments~\cite{Bonnici2013}.

\paragraph*{Graemlin}
The Graemlin data collection consists of medium sized and large graphs from microbial
networks.
The ten target graphs have between \numprint{1081} and \numprint{6726} vertices.
There are 420 pattern graphs with 4, 8, 16, 32, 64, 128, and 256 edges 
that are grouped into dense, semi-dense and sparse groups.
There are different versions of this data collection. One is labeled with unique labels,
while the others are labeled with 32, 64, 128, 256, 512, \numprint{1024} and
\numprint{2048} different
labels using a uniform distribution.
The low label versions especially have a high percentage of unsolved instances
and longer running times~\cite{Bonnici2013}. We use the 32-label version
since it is hard and still contains many solvable instances. We refer
to it as GRAEMLIN32.


%
%
%
%
\ifFull
\subsection{Metrics}
In order to gain insight into different facets of the algorithms' behavior 
we want to measure different aspects of the algorithm.

\paragraph*{Preprocessing time}
The time required for creating the node ordering and, in the case of RI-DS, the
initial domain assignment. 
This is in contrast to Bonnici~\etal where the domain assignment time is
included in the matching time. Because our parallelization only affects the
actual search it makes more sense to split the linear processing from the
parallel processing part.

\paragraph*{Matching time}
The time required for searching the state space representation. This includes
the time to start worker threads but does not include the preprocessing.

\paragraph*{Total time}
The time required for preprocessing an instance and subsequently solving it.
This does not include the time required to load the instance from disk.

\paragraph*{Search space size}
The number of pattern node and target node pairs considered during the search
process. This shows the effects on search space size of our improvements
to RI-DS\@.

\paragraph*{Match count}
The number of matches found for a certain instance. We compare all match counts
against the results of the reference implementation of RI 3.6 and RI-DS 3.51 to
ensure the validity of our results.


\paragraph*{Steal attempts}
The number of steal attempts by any individual worker.
We record both failed and successful steal attempts.

\paragraph*{Peak memory usage}
The peak memory usage reached while solving an instance.
\fi

\subsection{Experimental platform and results}
\ifFull
\subsection{Experimental setup}
We measure the running times with C++11's \texttt{chrono::high\_resolution\_timer}. 
Memory consumption is recorded using GNU
time\footnote{https://www.gnu.org/software/time/}. Search space size and steal
attempts are recorded by the workers and reported at the end of a run.
\fi

We run our experiments on a dedicated machine with 256 GB RAM and 2$\times$8
Intel\textregistered~Xeon\textregistered~E5\-2680 cores, running 
openSUSE 13.1 with Linux version 3.12.62\-52 (x86\_64).
Our code was compiled using GCC 4.8.1 (gcc-4\_8-branch revision 202388) and
optimization flag \texttt{-O3}. We use
TCMALLOC\footnote{\texttt{https://github.com/gperftools/gperftools/}} for memory
allocation and \texttt{pthread} for threading.

\ifFull
\subsection{Experimental results}
\fi 

In the detailed comparison of the algorithms below, 
we are chiefly concerned with the \emph{matching time} of the algorithms---the time it takes to enumerate all isomorphic subgraphs.
To measure the efficiency of our parallel implementation, we report the speedup of our algorithm as we increase the number of workers.  
One challenge to measuring the speedup is that the data collections contain many more short running instances than long running instances.
Since our goal is to solve long running instances quickly, we must prevent short running instances (which have little speedup) from dominating our measurements.
First, we compute the speedup as an arithmetic mean over the total runtime to process \emph{all instances} of a particular data collection (avg in our tables).
We report the standard error of the mean with red bars in our point plots.
We further compute the geometric mean (gmean in our tables) of the speedup of each instance. 
Moreover, we split instances into short and long running groups either by the time
required in the original RI/RI-DS implementation or against a single worker of our parallel implementation, depending on the comparison.

\ifFull
First we have a few small experiments that motivate different design aspects of
our algorithm.
Secondly we determine the right parameters for task granularity in the parallel implementation.
Thirdly we aim to understand how well our proposed improvements to RI-DS work
at reducing search space size.
Finally we compare the best version of our implementation against the
original RI and RI-DS implementations and analyze how well parallelization works 
for our problem.
\fi

\ifFull
\begin{figure}[h]
\floatsetup{capposition = below, floatrowsep =qquad,}
\begin{subfloatrow}
        \includegraphics[width=0.5\textwidth]{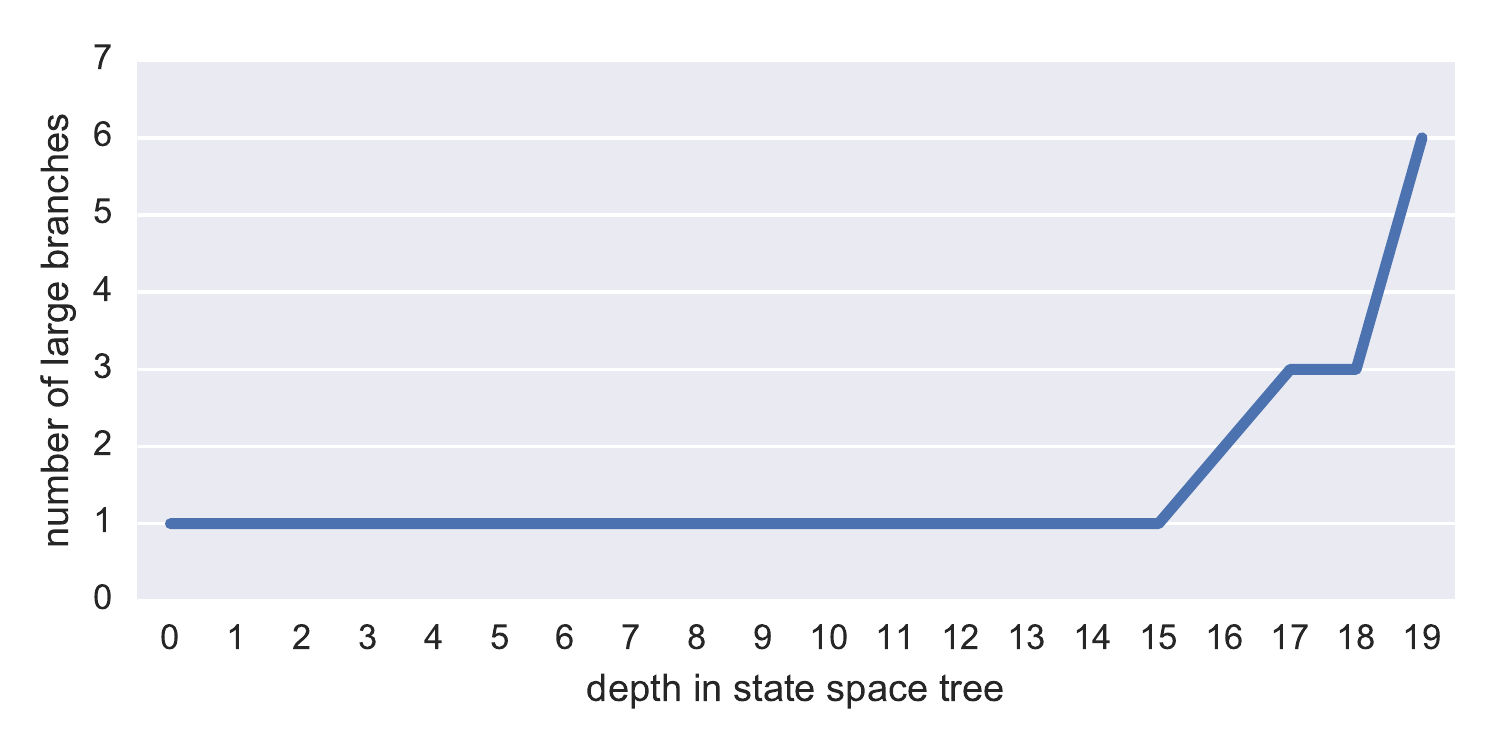}%
        \includegraphics[width=0.5\textwidth]{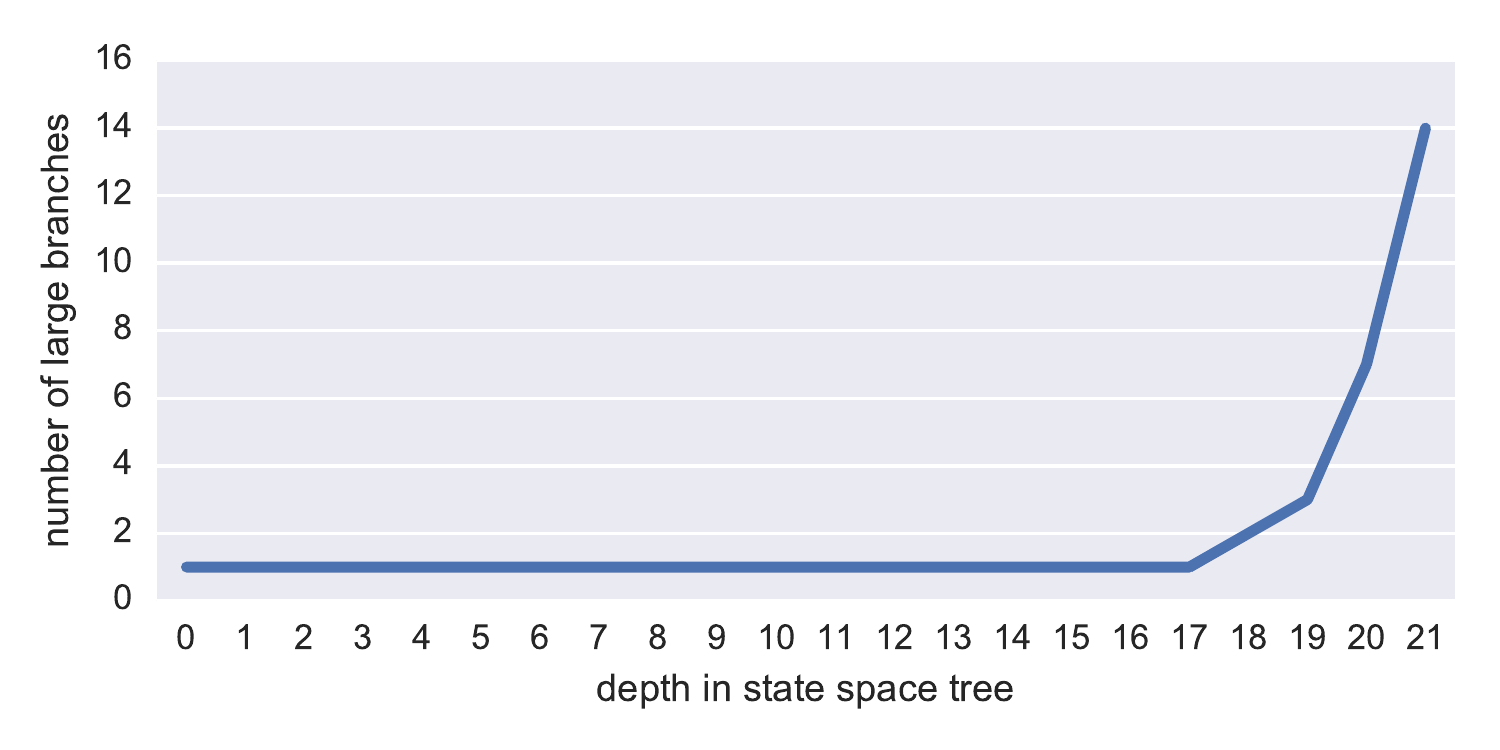}%
\end{subfloatrow}
\begin{subfloatrow}
        \includegraphics[width=0.5\textwidth]{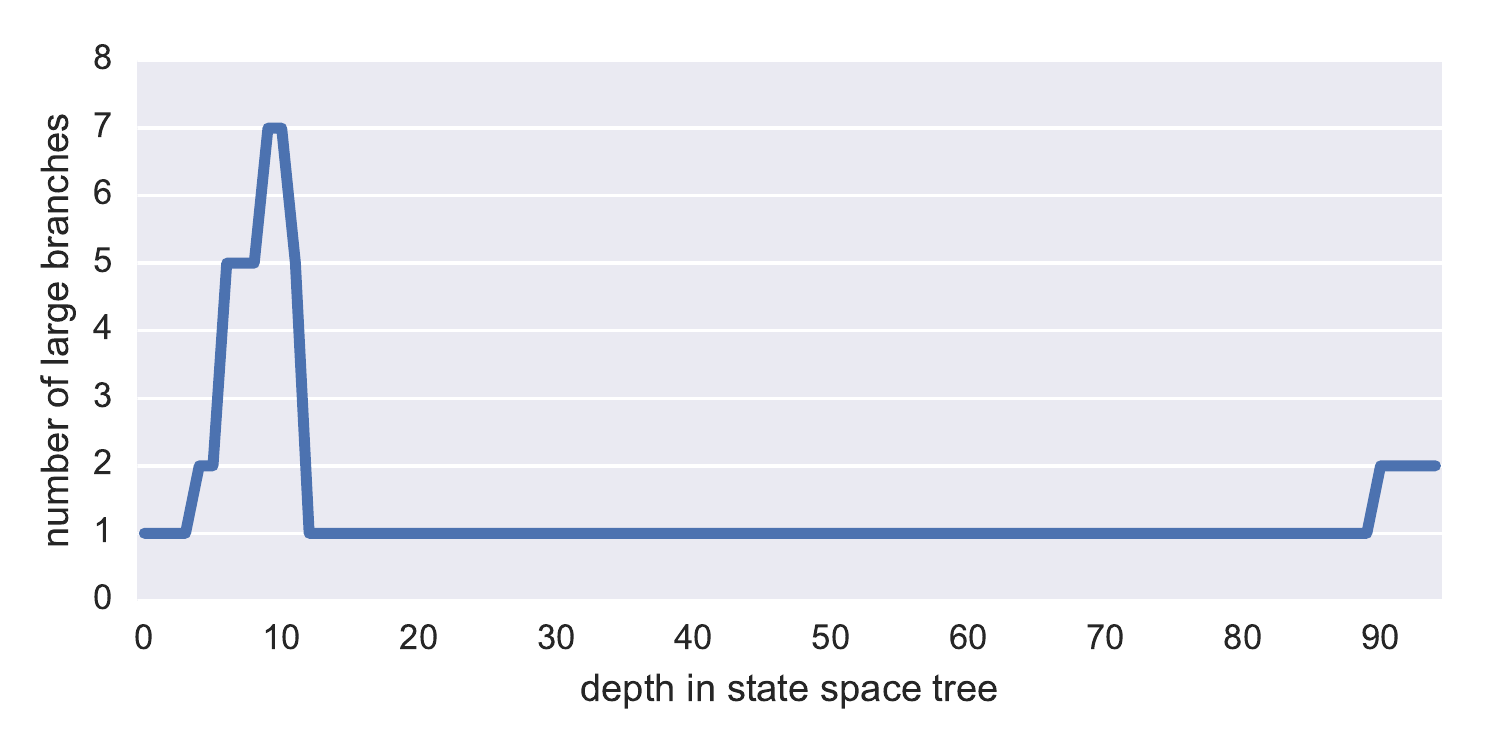}%
        \includegraphics[width=0.5\textwidth]{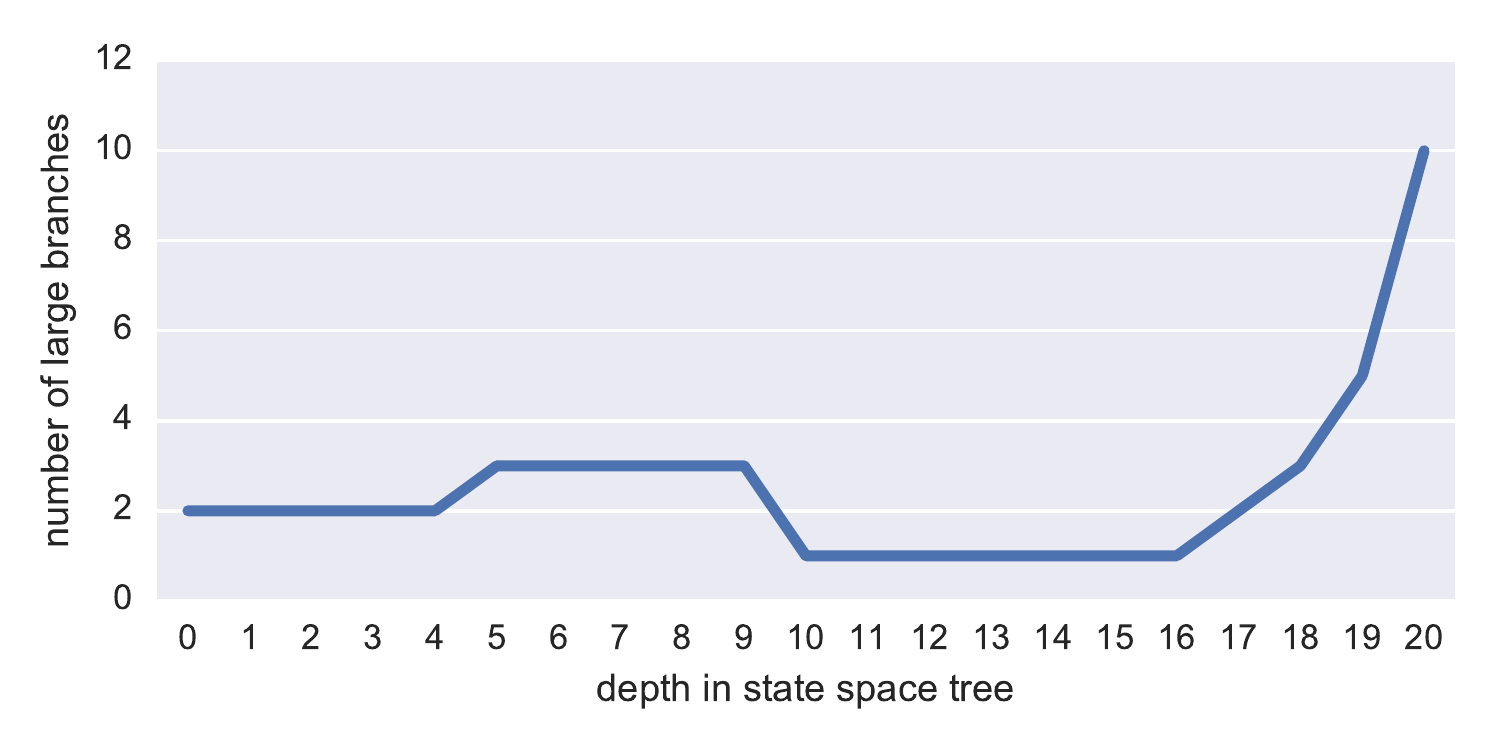}%
\end{subfloatrow}
\caption{Number of branches of the search space with at least 5\% of total
search space below them for every depth on four instances of PPI.}
\label{fig:search_space_parallel}
\end{figure}

\subsection{Search space parallelism}
In \fig~\ref{fig:search_space_parallel} we show the search space of four
sample instances (target graphs from PPI are \textsc{Homo\_sapiens.gfd},
\textsc{Drosophila\_melanogaster.gfd},
\textsc{Rattus\_norvegicus.gfd} and \textsc{Saccaromyces\_cerevisiae.gfd}). 
We see for every depth of the search space, the number of large branches, that
is branches with at least 5\% of the total search space below them. We
can see that, while there may at times be limited parallelism, for the majority
of the states to be explored at least some parallelism can be found. 
It should be noted that having phases with just a single large branch followed
by an increase in large branches is not necessarily a problem. A single
worker may only need to explore a few states to reach a depth where the number of
large branches increases.

These results were obtained by instrumenting the original implementation of RI
to record all states visited during the search process. They can of course only
serve to give some intuition as to the structure of the search space.
\fi

\ifSuggestedImage
\begin{figure}[t]
    \includegraphics{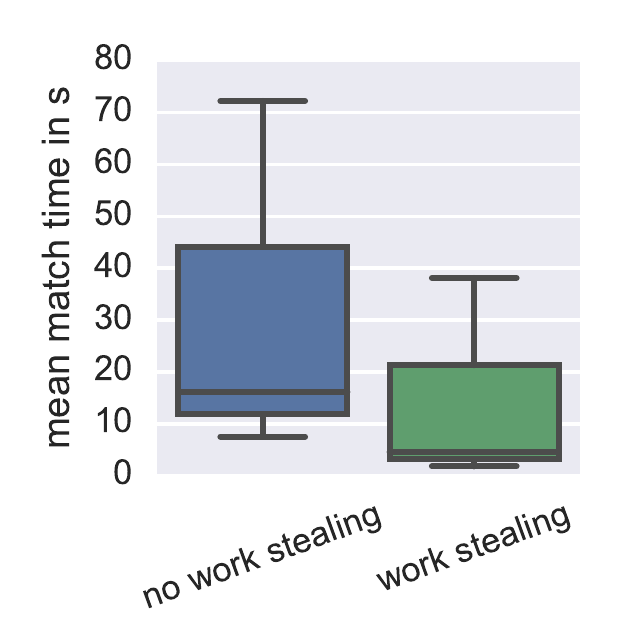}
    \includegraphics{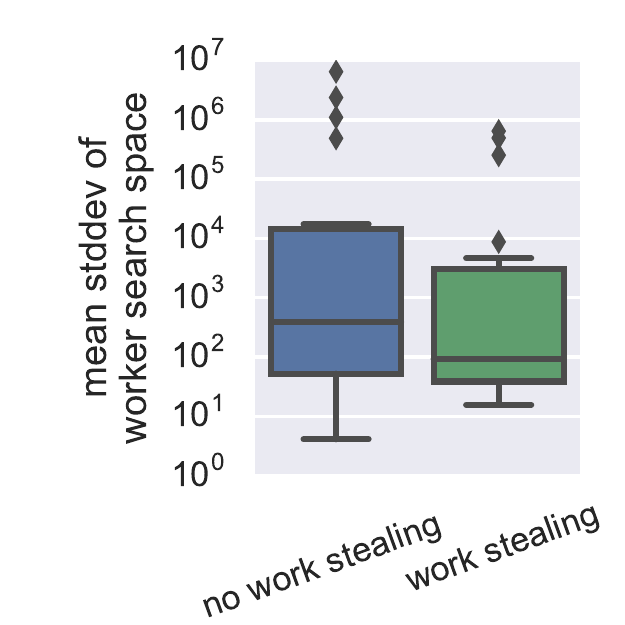}
\caption{The effects of work stealing with 16 workers on a random sample of instances from PPIS32.}
\label{fig:load_balancing}
\end{figure}

\else
\begin{figure}[]
    \includegraphics[width=\textwidth]{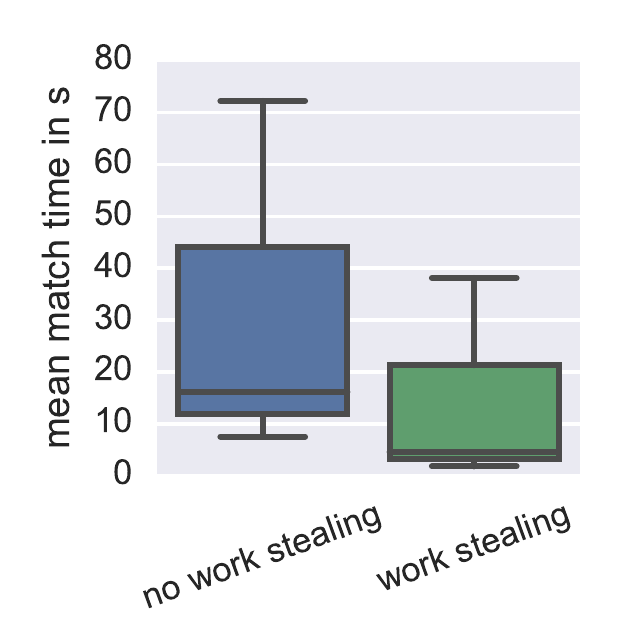}
    \caption{Effects of load balancing on match time with 16 threads on a
    random sample of long running (match time $>$ 1 second) instances of PPIS32.
    \imagenote{For this figure and \fig~\ref{fig:load_balancing_std}: box plots, "work stealing" "no work stealing", put side by side in one column in the paper. Squeeze horizontally, bars should be different colors. Same for Figure}
    }
\label{fig:load_balancing}
\end{figure}

\begin{figure}[]
    \includegraphics[width=\textwidth]{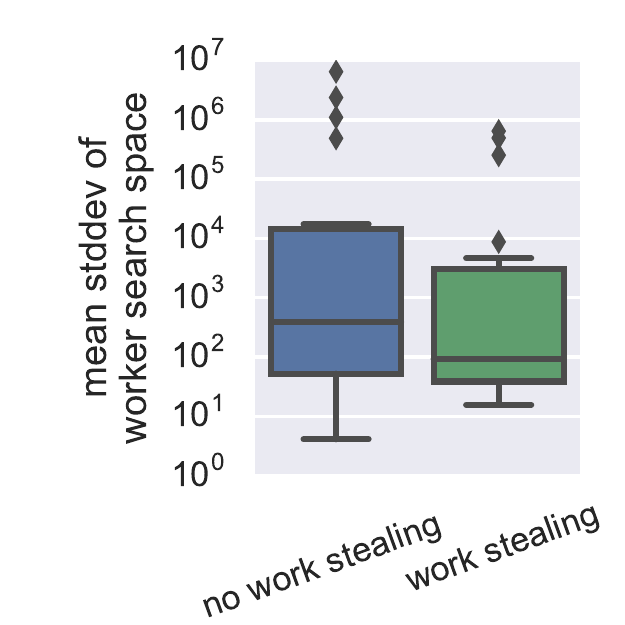}
    \caption{Effects of load balancing on distribution of work amongst 16 threads on a
    random sample of long running (match time $>$ 1 second) instances of PPIS32.}
\label{fig:load_balancing_std}
\end{figure}
\fi

\subsubsection{Work stealing}\label{ex:load_balancing}
We show the effect of work stealing on our parallelization by measuring the matching time for a sample of PPI with and without work stealing.
In \fig~\ref{fig:load_balancing}, we see that work stealing reduces the
time required to find a solution for 16 workers by a factor of 1.65.
Without work stealing, the number of states explored by all workers has a high standard deviation, indicating that work is unevenly distributed.

\ifSuggestedImage
\begin{figure*}[!h]
\centering
        \includegraphics[width=.22\textwidth]{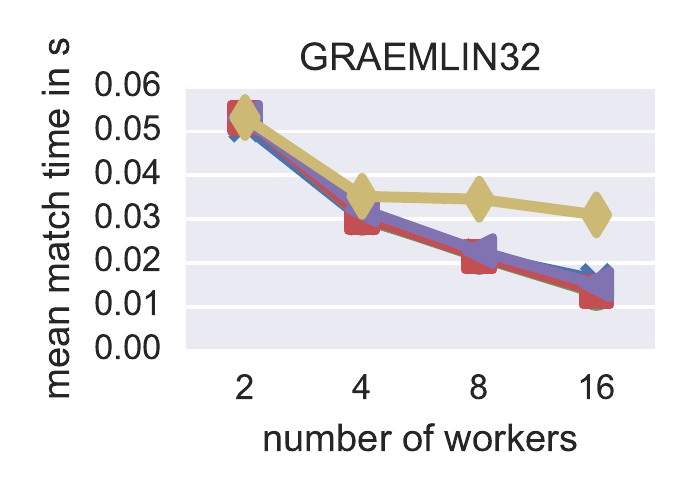}
        \includegraphics[width=.22\textwidth]{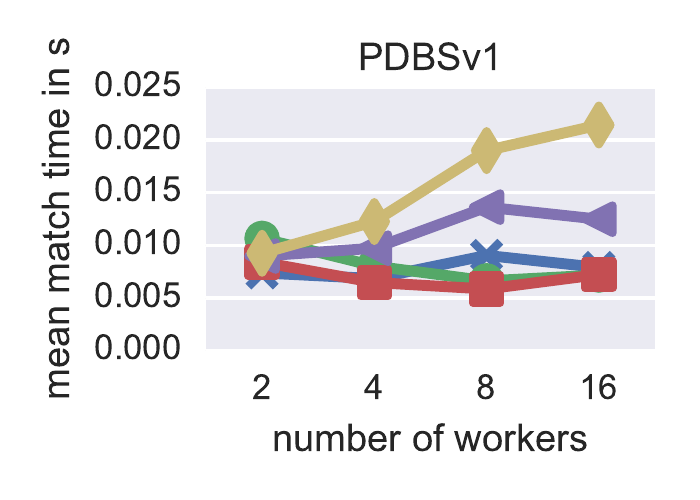}
        \includegraphics[width=.22\textwidth]{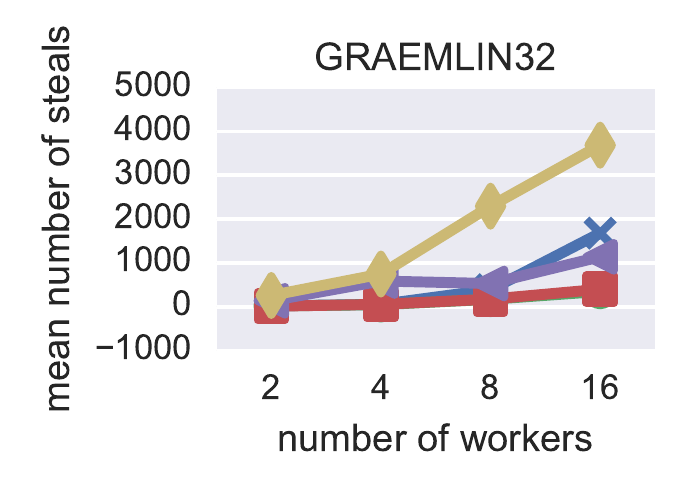}
        \includegraphics[width=.22\textwidth]{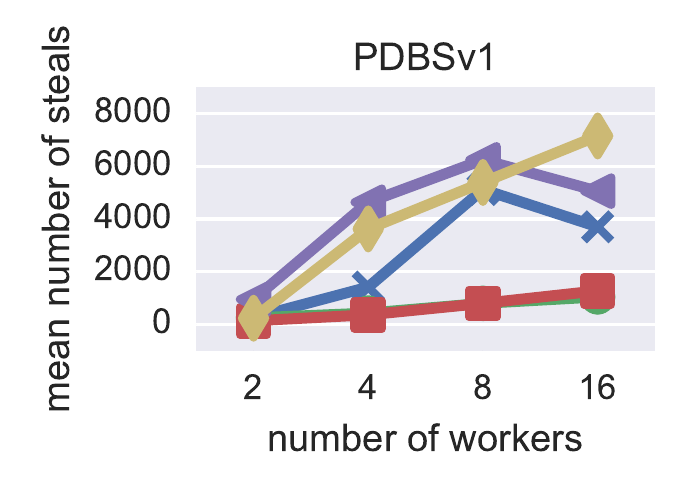}
        \includegraphics[width=.08\textwidth]{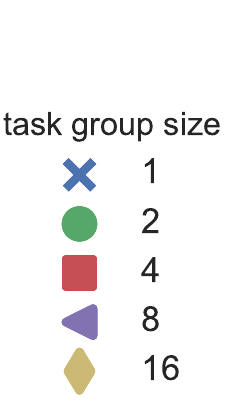}
    \caption{Effects of task group size on match time and number of steals.}
    \label{fig:task_group_size}
\end{figure*}

\else
\begin{figure}[]
\includegraphics[width=\textwidth]{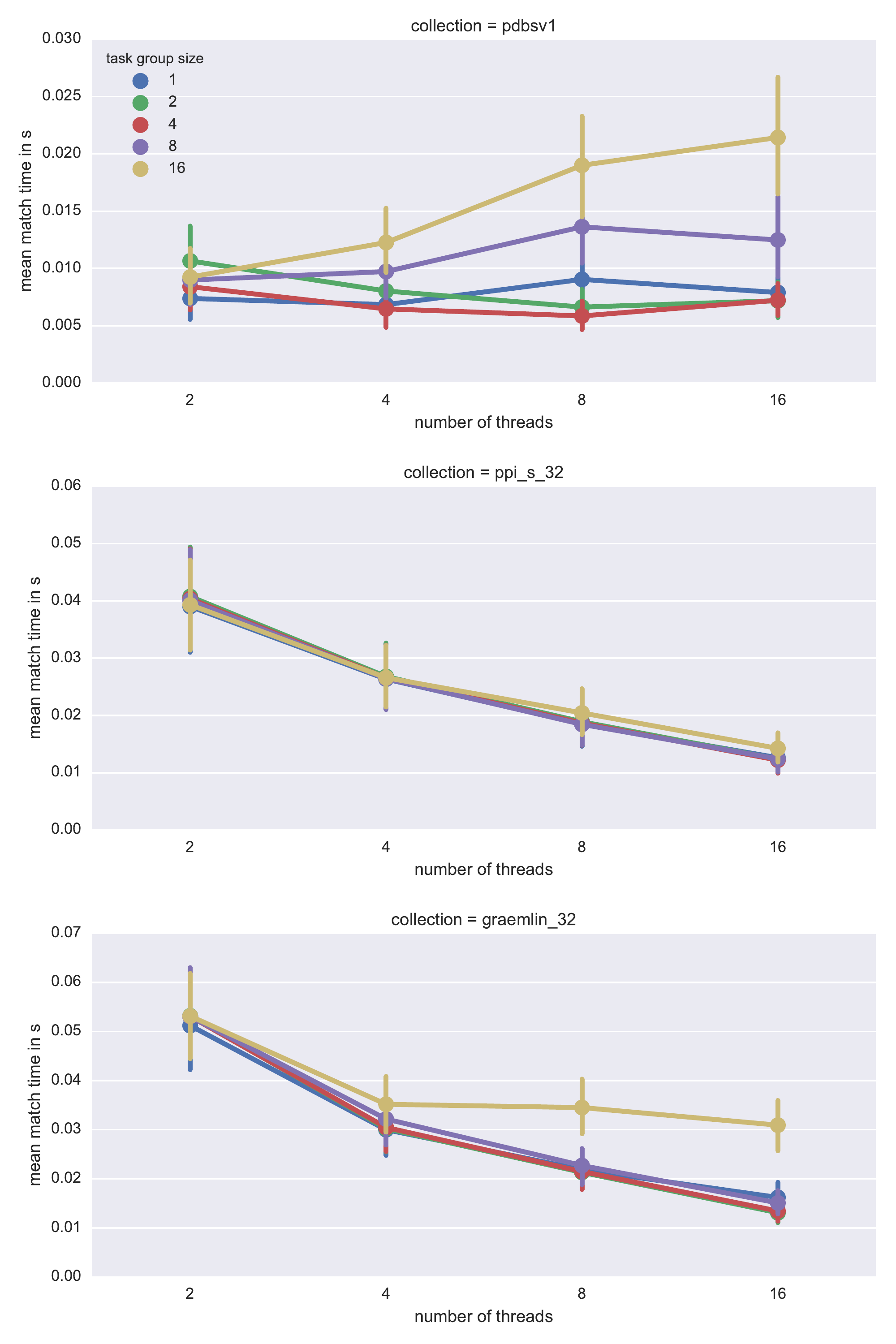}
\caption{The effects of task group size on match time.
\imagenote{For this figure and \fig~\ref{fig:task_group_size_steal}: Compress horizontally. Make 6 separate plot pdfs. Want 4 plots to span across top of page (over both columns).
Different symbol for each mark (triangle, square, circle, cross). or line style (dotted, dashed, etc) (not both). Remove "collection=".
Name data sets same as in paper: GREAMLIN32, etc. Use "number of workers" instead of "number of threads"}
}
\label{fig:task_group_size_time}
\end{figure}

\begin{figure}[]
\includegraphics[width=\textwidth]{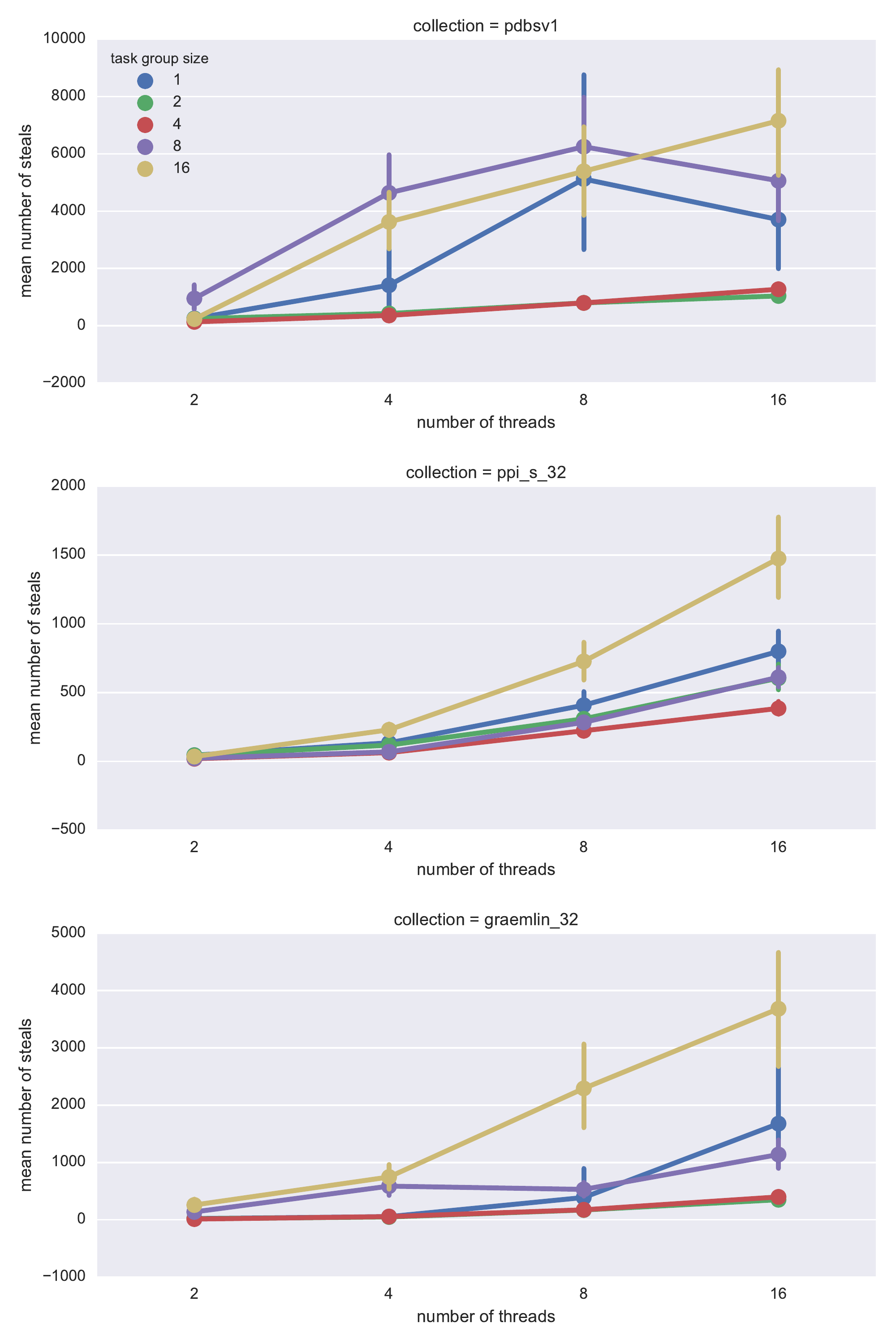}
\caption{The effects of task group size on number of steals performed.}
\label{fig:task_group_size_steal}
\end{figure}
\fi

\subsubsection{Task coalescing}
To measure the effect of task coalescing, we vary the task group size.
We experiment on a sample of short running instances of PPIS32, PDBSv1
and GRAEMLIN32 using 2, 4, 8, and 16 workers. We repeat each run 15
times to reduce variability and experiment with task group sizes of 1,
2, 4, 8 and 16 tasks. Results with a single worker are not included, as
larger task group sizes obviously reduce overhead in that case.

In \fig~\ref{fig:task_group_size} we can see that for PDBSv1 a
task group size of four is ideal.  For GRAEMLIN32 task group sizes of 
two and four yield the best running time for four or more workers. 
We note that PPIS32 (not shown here) has similar running times for all task group sizes,
except that size 16 is much worse.
Furthermore, the two rightmost plots in \fig~\ref{fig:task_group_size} clearly show that
less work stealing takes place for a task group size of two or four.
Notice that large task group sizes (16 for GRAEMLIN32 and 8-16 for PDBSv1) have many more steals than smaller task group sizes.
This behavior is due to the irregular search space:
few states (and few tasks) have much work below them in the state space tree. 
For large task group sizes, these tasks end up
in single task groups, which are stolen by a single worker. If multiple
such task groups are held by a single worker, then other workers can only steal small tasks, which finish quickly and lead to more steals.
%
For our remaining experiments, we use task group size four.



\subsubsection{Performance of parallel RI}\label{ex:performance_parallel_ri}
We measure the performance of parallel RI on PDBSv1 with 1, 2, 4, 8 and 16
workers. 
Like Bonnici et al.~\cite{Bonnici2013} we do not include RI-DS in the
comparison on PDBSv1 because it was designed for dense graphs. Likewise we do
not test parallel RI on the dense GRAEMLIN32 and PPIS32 sets but rather the
versions designed for dense graphs: parallel RI-DS and our improvements.

\begin{figure}[h]
\centering\includegraphics[width=0.75\textwidth]{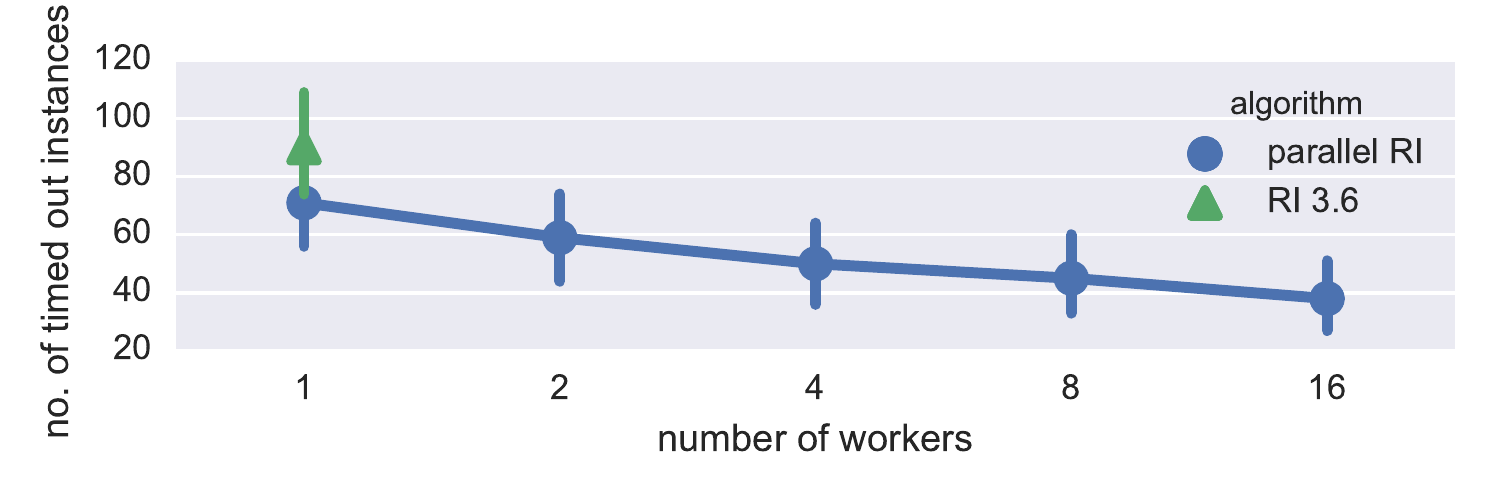}
\caption{Number of timed out instances on PDBSv1 for RI 3.6 and parallel RI.}
\label{fig:ri_pdbsv1_timed_out}
\end{figure}

\ifFull
\begin{figure}[h]
\centering\includegraphics[width=0.75\textwidth]{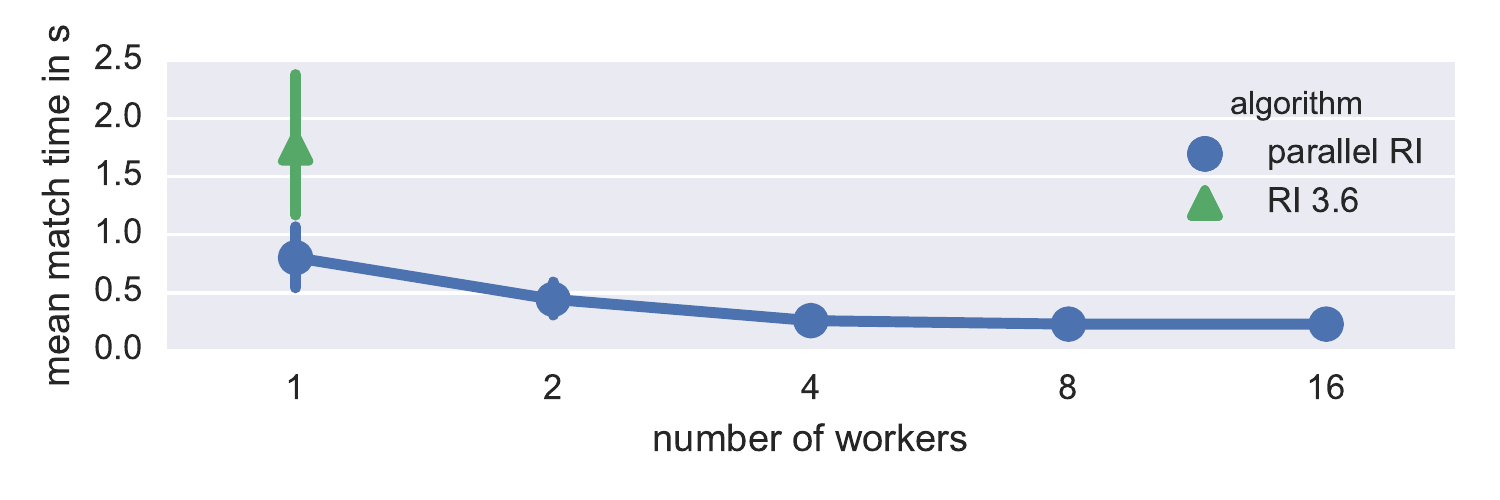}
\caption{Match time on PDBSv1 for RI 3.6 and parallel RI.}
\label{fig:ri_pdbsv1_match_time}
\end{figure}

\begin{figure}[h]

    \centering{\includegraphics[width=0.75\textwidth]{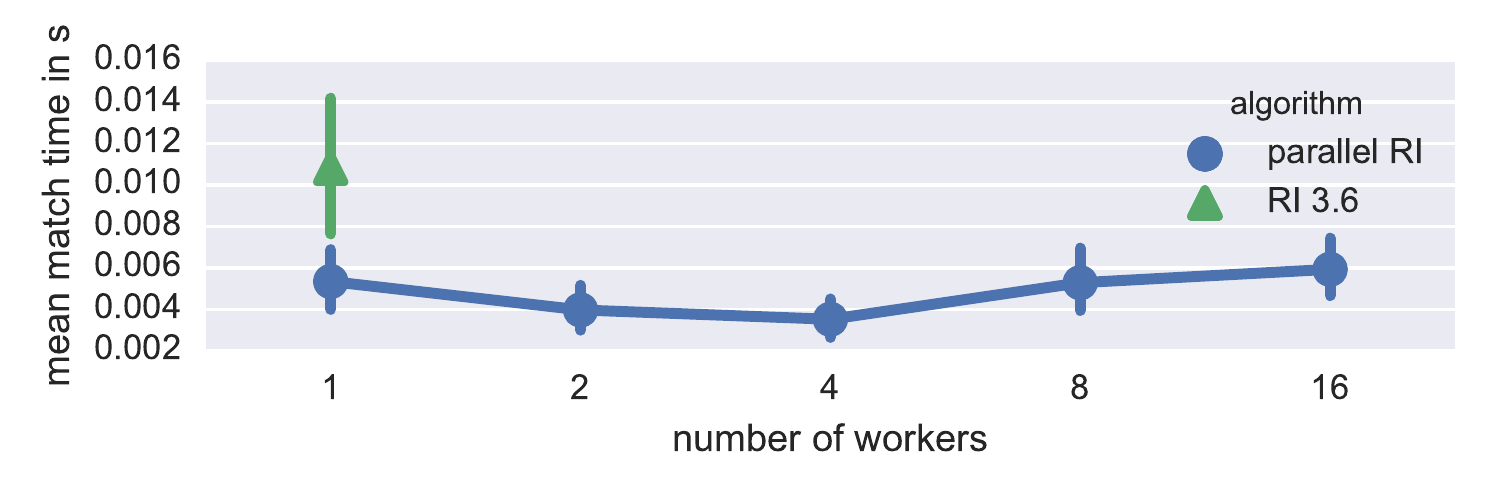}}
\caption{Match time on PDBSv1 for RI 3.6 and parallel RI on instances with a
match time of less than one second.}
\label{fig:ri_pdbsv1_match_time_short}
\end{figure}
\fi

\begin{figure}[h]
\centering\includegraphics[width=0.75\textwidth]{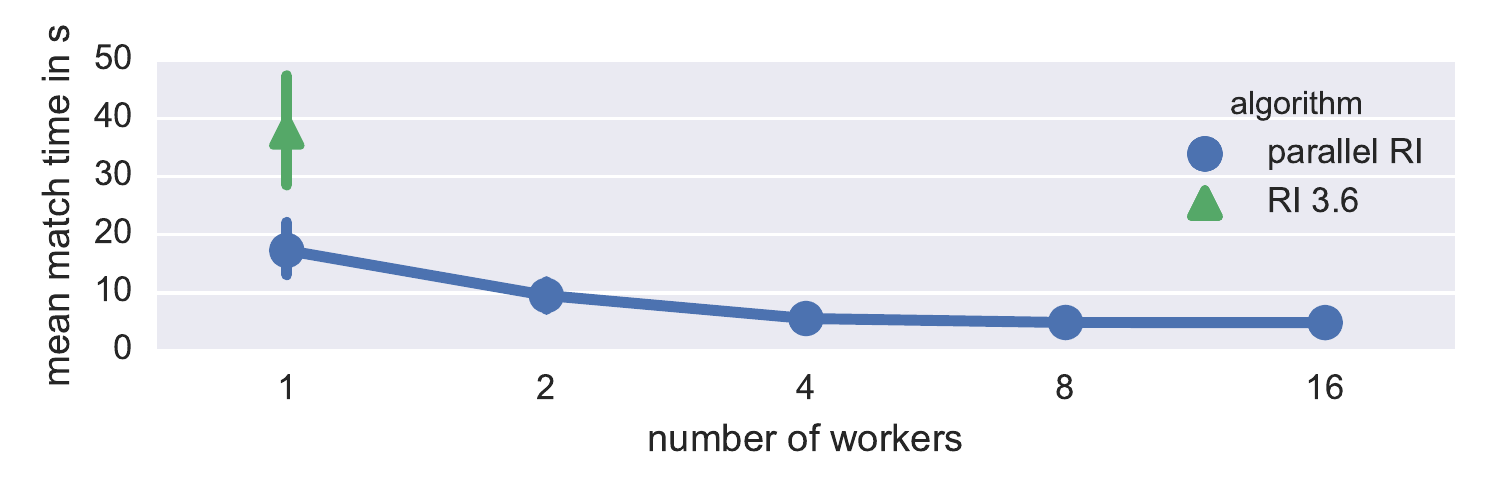}
\caption{Match time on PDBSv1 for parallel RI on instances with a
match time of more than one second.}
\label{fig:ri_pdbsv1_match_time_long}
\end{figure}

\ifFull
\begin{figure}[h]
\centering\includegraphics[width=\textwidth]{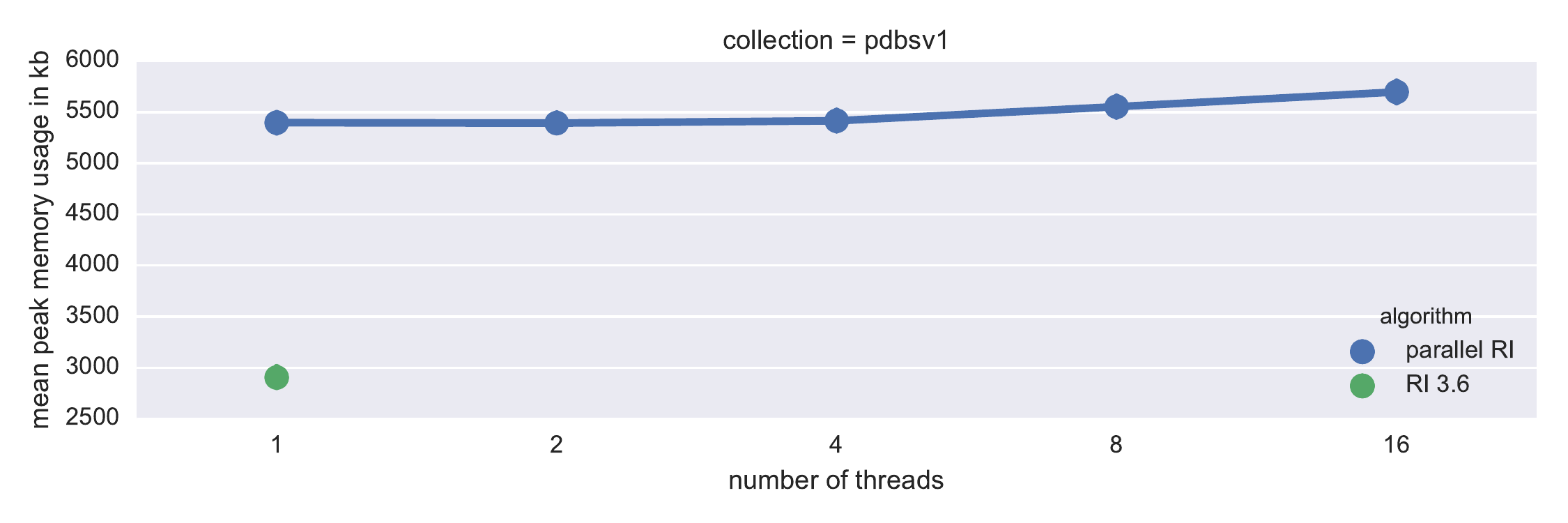}
\caption{Memory usage on PDBSv1 for RI 3.6 and parallel RI.}
\label{fig:ri_pdbsv1_memory}
\end{figure}
\fi

\ifOurBaseline
\ifNewNumbers
\begin{table*}[]
   \small
   \centering
   \caption{Speedup of parallel RI over our implementation with one worker,  for all, short, and long instances. Maximum speedups marked with a $*$ exceeded the worker count, due to a lack of precision when timing short running instances.}\label{tab:ri_speedup_ourbaseline}
   \begin{tabular}{c r r r r r r r r r}
      \toprule
      \# workers & \multicolumn{3}{c }{all instances} & \multicolumn{3}{c }{short ($<1$ sec.)} & \multicolumn{3}{c }{long ($\ge 1$ sec.)}\\
      \cmidrule(rr){2-4}
      \cmidrule(rr){5-7}
      \cmidrule(rr){8-10}
      & avg & gmean & $\max$ & avg & gmean & $\max$ & avg & gmean & $\max$ \\
      \midrule
        2  & 1.78 & 1.20 & 3.37$^*$ & 1.71 & 1.17 & 3.37$^*$ & 1.78 & 1.75 & 1.91 \\
        4  & 3.03 & 1.31 & 5.20$^*$ & 2.30 & 1.25 & 5.20$^*$ & 3.03 & 2.88 & 3.53 \\
        8  & 4.28 & 0.80 & 5.92\phantom{$^*$} & 2.12 & 0.73 & 5.47\phantom{$^*$} & 4.29 & 3.97 & 5.92 \\
        16  & 5.90 & 0.64 & 10.02\phantom{$^*$} & 1.84 & 0.57 & 5.23\phantom{$^*$} & 5.96 & 5.12 & 10.02 \\
      \bottomrule
   \end{tabular}
\end{table*}

\else
\ifIncludeMin
\begin{table*}[]
   \centering
   \caption{Speedup of parallel RI over our implementation with one worker.}\label{tab:ri_speedup_ourbaseline}
   \begin{tabular}{c r c l r c l r c l}
      \toprule
      \# workers & \multicolumn{3}{c }{all instances} & \multicolumn{3}{c }{short ($<1$ sec.)} & \multicolumn{3}{c }{long ($\ge 1$ sec.)}\\
      \cmidrule(lr){2-4}
      \cmidrule(lr){5-7}
      \cmidrule(lr){8-10}
      & avg & $\min$ & $\max$ & avg & $\min$ & $\max$ & avg & $\min$ & $\max$ \\
      \midrule
    2  & 1.89 & 0.35 & 2.18 & 1.43 & 0.35 & 2.18 & 1.89 & 0.85 & 2.04 \\
    4  & 3.34 & 0.29 & 3.89 & 1.73 & 0.29 & 3.28 & 3.35 & 1.70 & 3.89 \\
    8  & 4.29 & 0.03 & 6.22 & 1.15 & 0.03 & 3.60 & 4.33 & 1.08 & 6.22 \\
    16  & 4.34 & 0.00 & 8.91 & 1.07 & 0.00 & 2.99 & 4.39 & 0.68 & 8.91 \\
      \bottomrule
   \end{tabular}
\end{table*}
\else
\ifIncludeMax
\begin{table}[]
   \centering
   \caption{Speedup of parallel RI over itself with one worker on the PDBSv1 data collection, for all, short ($<1$ sec.), and long ($\geq1$ sec.) instances.}\label{tab:ri_speedup_ourbaseline}
   \begin{tabular}{c r l r l r l}
      \toprule
      \# workers & 
      \multicolumn{6}{c}{Speedup of parallel RI} \\
      & \multicolumn{2}{c }{all instances} & \multicolumn{2}{c }{short} & \multicolumn{2}{c }{long}\\
      \cmidrule(lr){2-3}
      \cmidrule(lr){4-5}
      \cmidrule(lr){6-7}
      & avg & $\max$ & avg & $\max$ & avg & $\max$ \\
      \midrule
    2  & 1.89 & 2.18 & 1.43 & 2.18 & 1.89 & 2.04 \\
    4  & 3.34 & 3.89 & 1.73 & 3.28 & 3.35 & 3.89 \\
    8  & 4.29 & 6.22 & 1.15 & 3.60 & 4.33 & 6.22 \\
    16  & 4.34 & 8.91 & 1.07 & 2.99 & 4.39 & 8.91 \\
      \bottomrule
   \end{tabular}
\end{table}
\else
\begin{table}[]
   \centering
   \caption{Speedup of parallel RI over itself with one worker on the PDBSv1 data collection, for all, short ($<1$ sec.), and long ($\geq1$ sec.) instances.}\label{tab:ri_speedup_ourbaseline}
   \begin{tabular}{c r r r}
      \toprule
      \# workers & 
      \multicolumn{3}{c}{Speedup of parallel RI} \\
      & \multicolumn{1}{c }{all} & \multicolumn{1}{c }{short ($<1$ sec.)} & \multicolumn{1}{c }{long ($\geq1$ sec.)}\\
      \cmidrule(lr){2-2}
      \cmidrule(lr){3-3}
      \cmidrule(lr){4-4}
      & avg & avg & avg \\
      \midrule
      2   & 1.89 & 1.43 & 1.89 \\
      4   & 3.34 & 1.73 & 3.35 \\
      8   & 4.29 & 1.15 & 4.33 \\
      16  & 4.34 & 1.07 & 4.39 \\
      \bottomrule
   \end{tabular}
\end{table}
\fi 
\fi 
\fi 

\ifFull
\begin{table}[]
\small
   \centering
   \caption{Speedup of parallel RI over RI 3.6 on the PDBSv1 data collection, for all, short ($<1$ sec.) and long ($\geq1$ sec.) instances.}\label{tab:speedup_ri_original}
   \begin{tabular}{r r r r}
      \toprule
      \# workers &
      \multicolumn{3}{c}{Speedup over RI version 3.6} \\
      \cmidrule(lr){2-4}
      & all instances & short ($<1$ sec.) & long ($\ge1$ sec.) \\

\midrule
1 & 2.18 & 2.02 & 2.18 \\
2 & 3.94 & 2.72 & 3.95 \\
 4 & 6.73 & 3.07 & 6.77 \\
 8 & 7.60 & 2.04 & 7.73 \\
16 & 7.61 & 1.82 & 7.75 \\
      \bottomrule
   \end{tabular}
\end{table}

\fi

As can be seen in \fig~\ref{fig:ri_pdbsv1_timed_out}, multithreading significantly
reduces the number of unsolved instances for long running instances. While RI 3.6
fails to solve 90 instances in the time limit, parallel RI leaves only 38 instances unsolved
with 16 workers.
\ifFull
For our time comparison for parallel RI\ifFull in \fig~\ref{fig:ri_pdbsv1_match_time}\fi, we
only include instances solved by RI 3.6 within the time limit. We do this to
prevent skewing the results towards the algorithm that solves the fewest
hard instances.
\fi
Surprisingly, RI 3.6 is
slower than parallel RI with a single worker, which we believe
is due to subtle implementation differences.
Since it is faster, we therefore compare our parallel RI implementation to itself with one worker. However, before this comparison,
we briefly mention how parallel RI compares to RI 3.6. On instances solved by RI 3.6 within the time limit, parallel RI with 16 threads beats RI 3.6 by an average factor of 3.07 on short running instances with 4 workers, and
a speedup of 7.75 with 16 workers on long running instances.
\ifFull
We can see in \fig~\ref{fig:ri_pdbsv1_match_time} that most of the benefits
of parallelization occur with up to four workers.
This is likely caused by the large number of
easy to solve instances in PDBSv1.
\else

Of the \numprint{1760} query graphs in the PDBSv1 data collection,
only 160 have a match time of longer than one second
with our single threaded implementation. Since short instances do not benefit
much from 
\fi
\ifFull
parallelism, we only briefly discuss them.
\fig~\ref{fig:ri_pdbsv1_match_time_short} we can see that for short running
instances using more than four workers actually results in an increase in match
time. %
\else
parallelism (as can be seen in Table~\ref{tab:ri_speedup_ourbaseline}), 
on average, using more than four workers increases match time. %
\fi
This is not surprising, as easily solved instances 
have a smaller search space than hard ones, offering little opportunity
for improved performance with parallelism. This behavior mirrors 
the slowdown observed by Acar et al.~\cite{Acar2013}---where private
deque based work stealing reduced search space parallelism.
However, for long running instances, increasing the number of workers always
improves the speedup. As seen in Table~\ref{tab:ri_speedup_ourbaseline},
parallel RI reaches an average speedup of 5.96 when running with 16 workers--and even achieves
a maximum speedup of 10.01 on some long running instances.
This is further corroborated in \fig~\ref{fig:ri_pdbsv1_match_time_long}, where we see that 
the average match time decreases as we increase the number of workers. The 
number of unsolved instances decreases similarly, as seen in \fig~\ref{fig:ri_pdbsv1_timed_out}.
We do, however, note that the number of unsolved instances does not decrease linearly with the number of workers. 
\ifIncludeMax
Since a higher speedup of up to 8.91 was achieved on some instances, 
we may assume that the limited speedup here is at least in part due to a lack of 
parallelism in the search space of most PDBSv1 instances.
\fi

\ifFull
The overall speedup compared to RI 3.6 is given in
Table~\ref{tab:speedup_ri_original}. 
On the
short running instances we achieve the best speedup of 3.07 using 4 threads and
on long running instances we achieve a speedup of 7.75 using 16 threads.
\else

\ifFull
In \fig~\ref{fig:ri_pdbsv1_memory} we can clearly see that RI 3.6 has much
lower memory consumption. The higher consumption of parallel RI compared to RI
3.6 is probably in part due to increased usage of the C++ standard library and
larger binary size (the binary size of RI 3.6 is 48,461 bytes compared to
118,544 bytes for the parallel implementation).
No efforts were undertaken to lower base memory consumption. We can see that
increases in parallelism result in a very moderate increase in memory usage.
\fi

\subsubsection{RI-DS-SI and RI-DS-SI-FC with domain size ordering and forward
checking}\label{ex:rids_improved}

\ifSuggestedImage
\begin{figure*}[t]
\centering
        \includegraphics[width=0.49\textwidth]{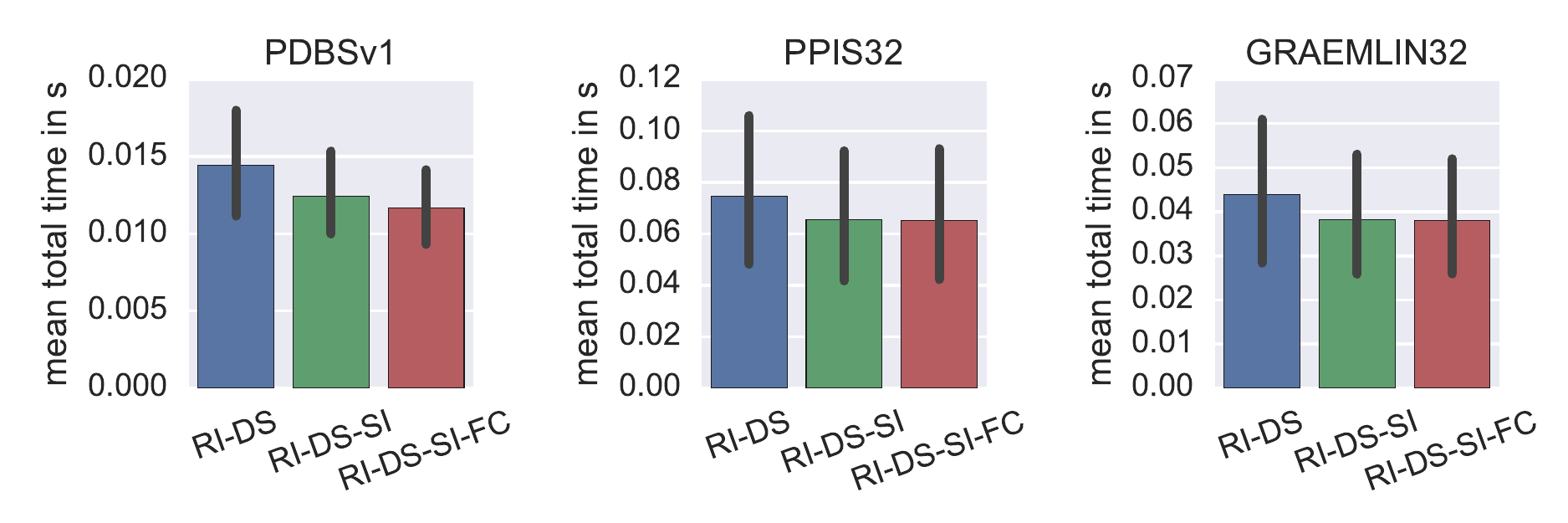}
        \includegraphics[width=0.49\textwidth]{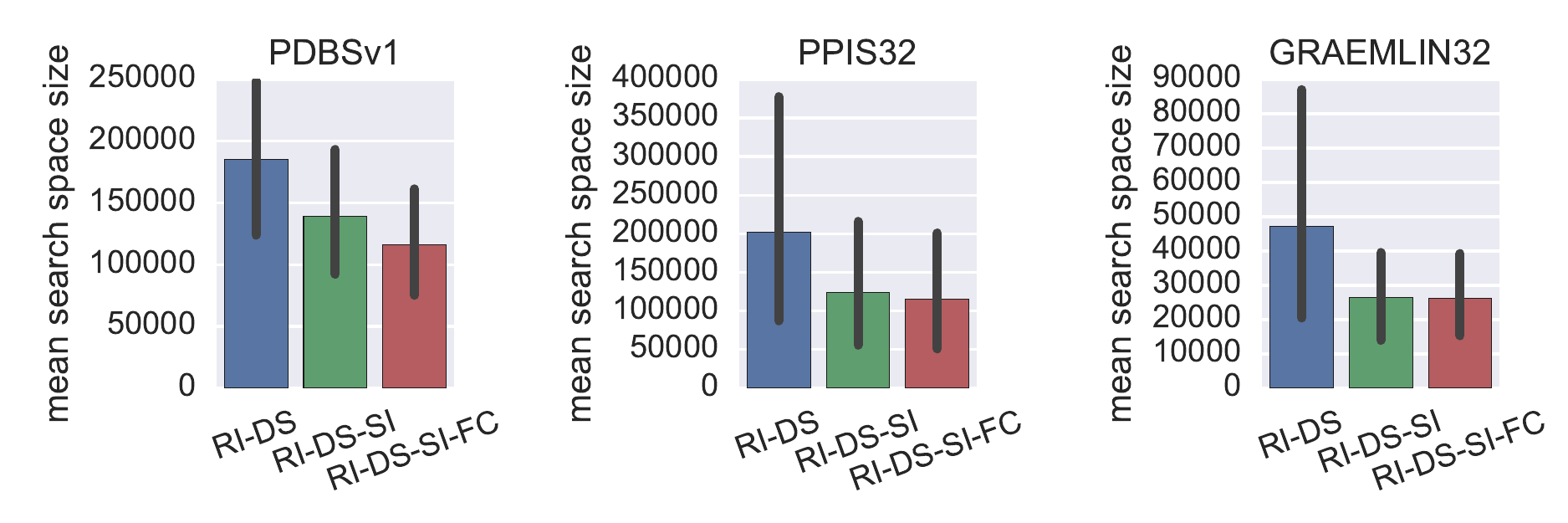}
    \caption{Search space reduction over, and single threaded run time
        comparison with, RI-DS. 
    }
    \label{fig:rids_improved_fast_search_space}
\end{figure*}

\else
\begin{figure}[]
\includegraphics[width=\textwidth]{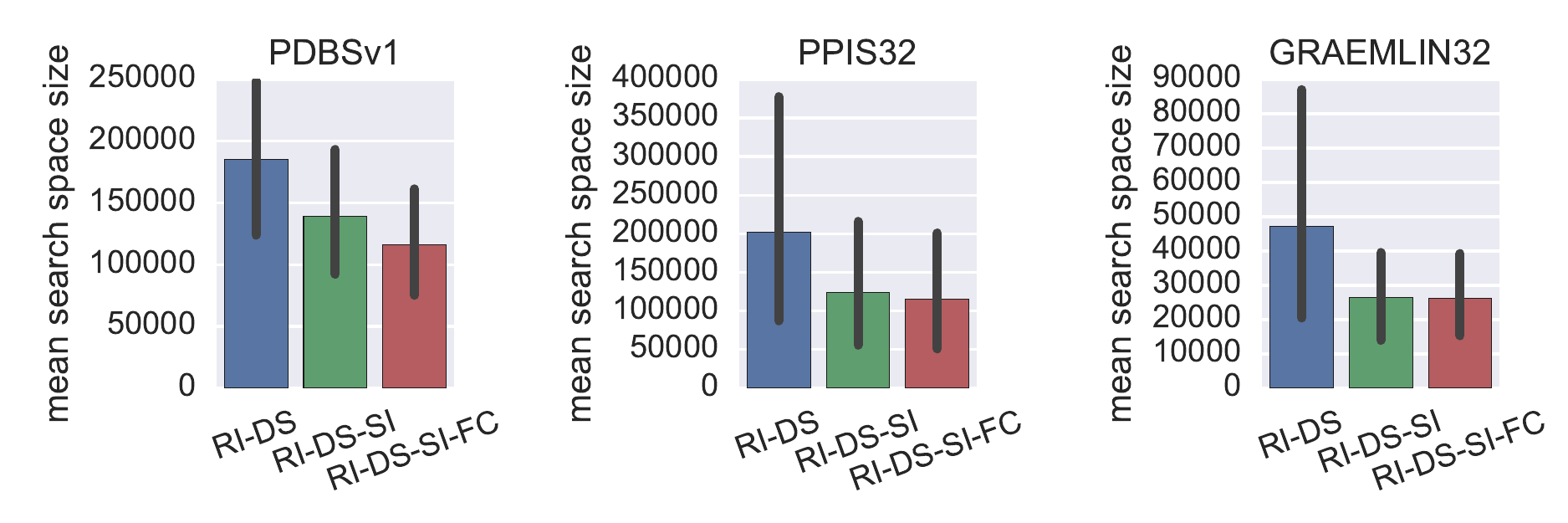}
\caption{Search space size on GRAEMLIN32, PPIS32 and PDBSv1 for our
implementation of RI-DS in single threaded mode with a time limit of one second
comparing different techniques.
\imagenote{For this figure and \fig~\ref{fig:rids_improved_fast_time}: box plots, compress horizontally so they fit across top of page? Not really
sure how to do that and still make readable. {\bf save for last!}}
}
\label{fig:rids_improved_fast_search_space}
\end{figure}

\begin{figure}[]
    \includegraphics[width=\textwidth]{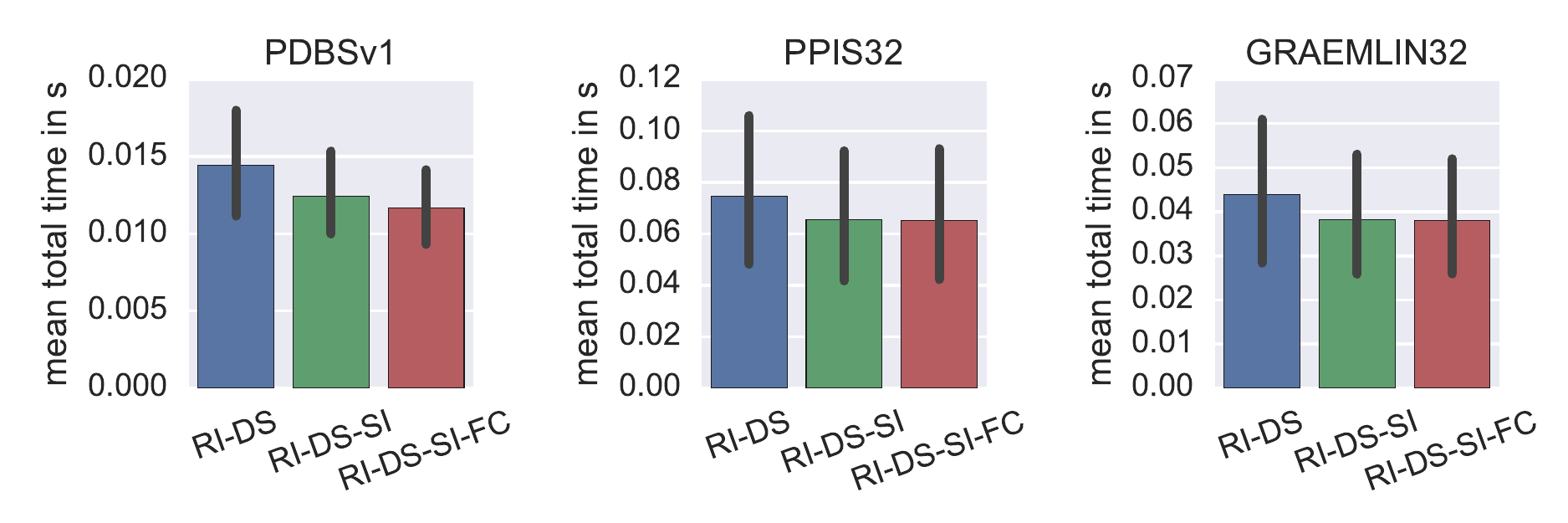}
\caption{Total time on GRAEMLIN32, PPIS32 and PDBSv1 for our implementation of
RI-DS in single threaded mode with a time limit of one second comparing
different techniques.}
\label{fig:rids_improved_fast_time}
\end{figure}
\fi

We measure the search space size and total time on PDBSv1, PPIS32 and
GRAEMLIN32 with and without our improvements (see
Section~\ref{sec:content:rids-improved}) on short running instances with a time
limit of one second and on a sample of longer running instances of PPIS32 and
GRAEMLIN32.

\paragraph*{Search space size}
In \fig~\ref{fig:rids_improved_fast_search_space} we give the search space
size and total time for domain size ordering (RI-DS-SI) and for domain size ordering with
forward checking (RI-DS-SI-FC). Domain size ordering clearly reduces 
search space size and total time for all data collections.
However, forward checking is less clear. On PDBSv1, forward checking
clearly reduces search space size as well as total time. 
For PPIS32 we see a slight improvement in search space size 
but no change in time, while for GRAEMLIN32 the search space size remains
essentially the same.

\begin{figure}[]
\floatsetup{capposition = below, floatrowsep =qquad,}
\ffigbox{%
\begin{subfloatrow}
    \ffigbox[0.5\textwidth]{\caption{Search space size.}}{%
        \includegraphics[width=0.35\textwidth]{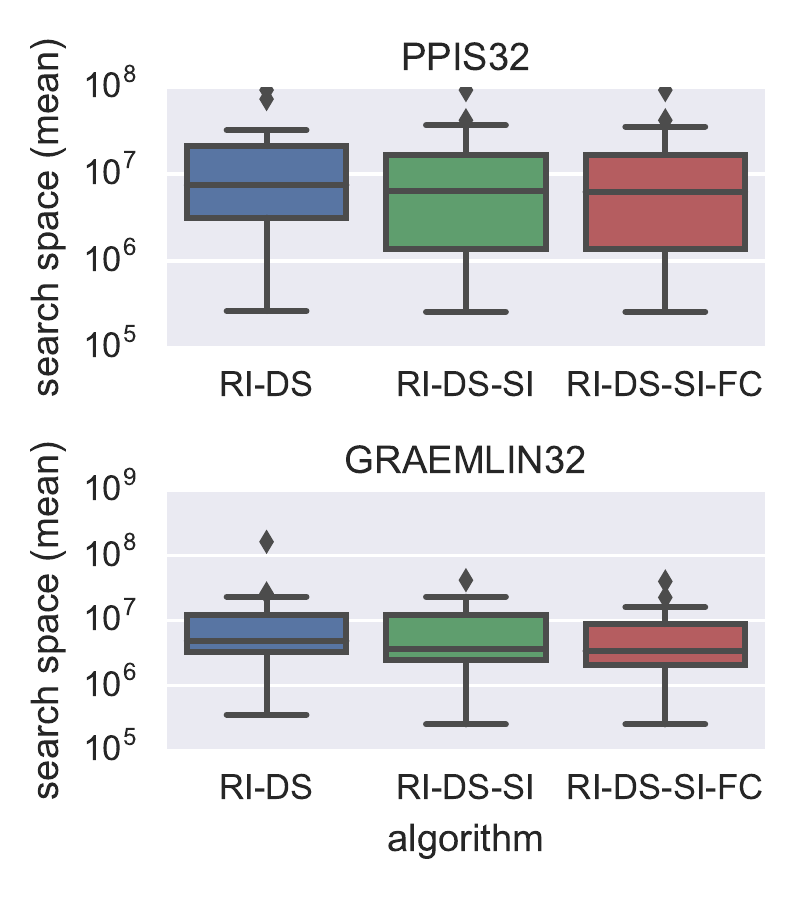}%
  }
    \ffigbox[0.5\textwidth]{\caption{Search speed.}}{%
    \includegraphics[width=0.35\textwidth]{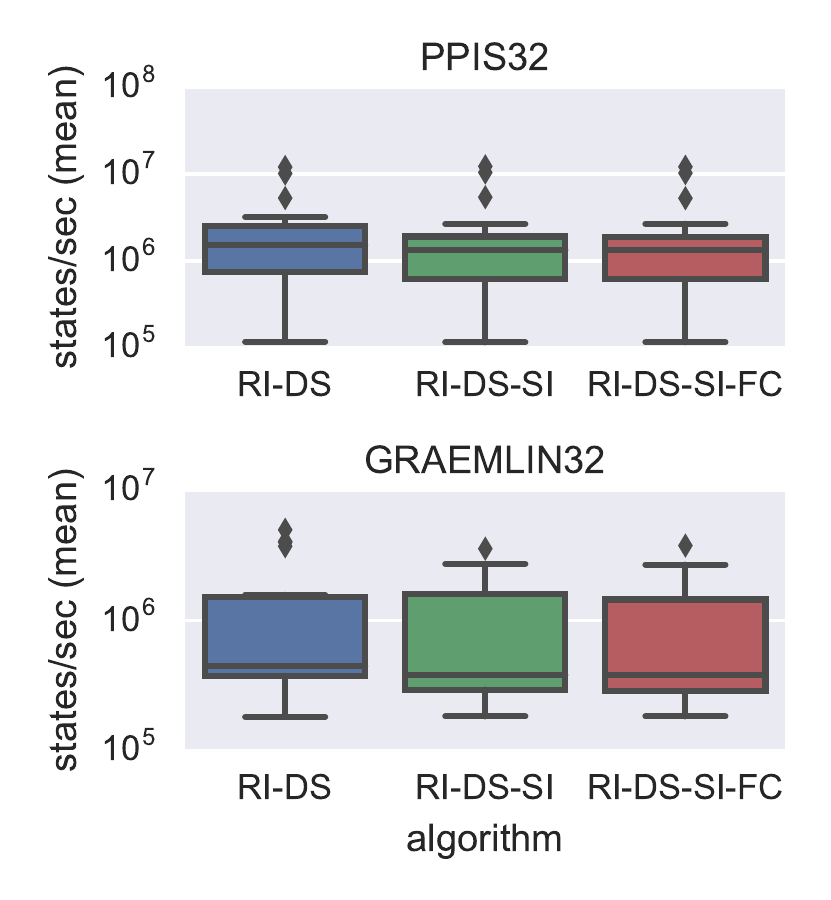}%
  }
\end{subfloatrow}
}{%
\caption{Search space on a random sample of long running instances from
GRAEMLIN32 and PPIS32 for our implementation of RI-DS and its variants with one worker.~\imagenote{box plot, larger font, ``states/sec (mean)''?}}
\label{fig:rids_improved_slow_search_space}
}
\end{figure}

\ifSuggestedImage
\begin{figure}[t]
\centering
        \centering
        \includegraphics[width=0.75\textwidth]{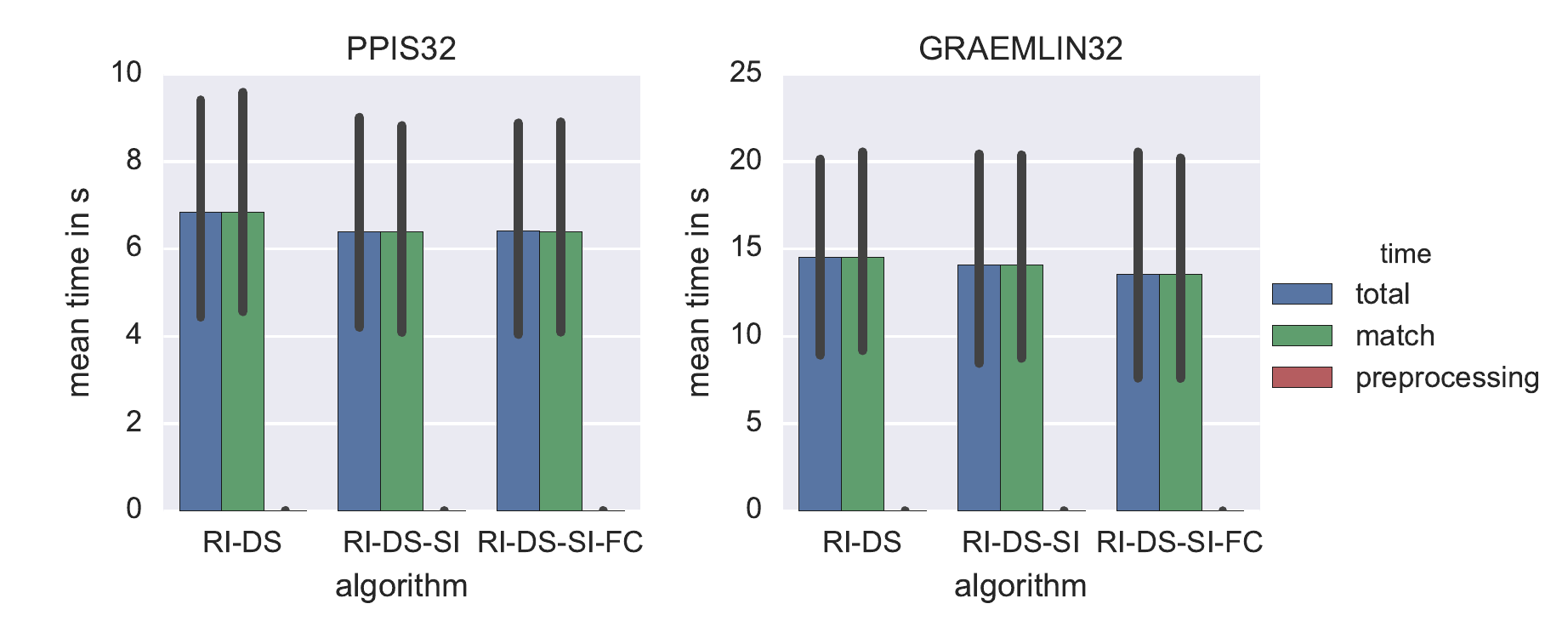}%
    \caption{The running time of our implementation of RI-DS and its variants. Note that preprocessing time is negligible.
    }
\label{fig:rids_improved_slow_times}
\end{figure}

\else
\begin{figure}[]
\includegraphics[width=\textwidth]{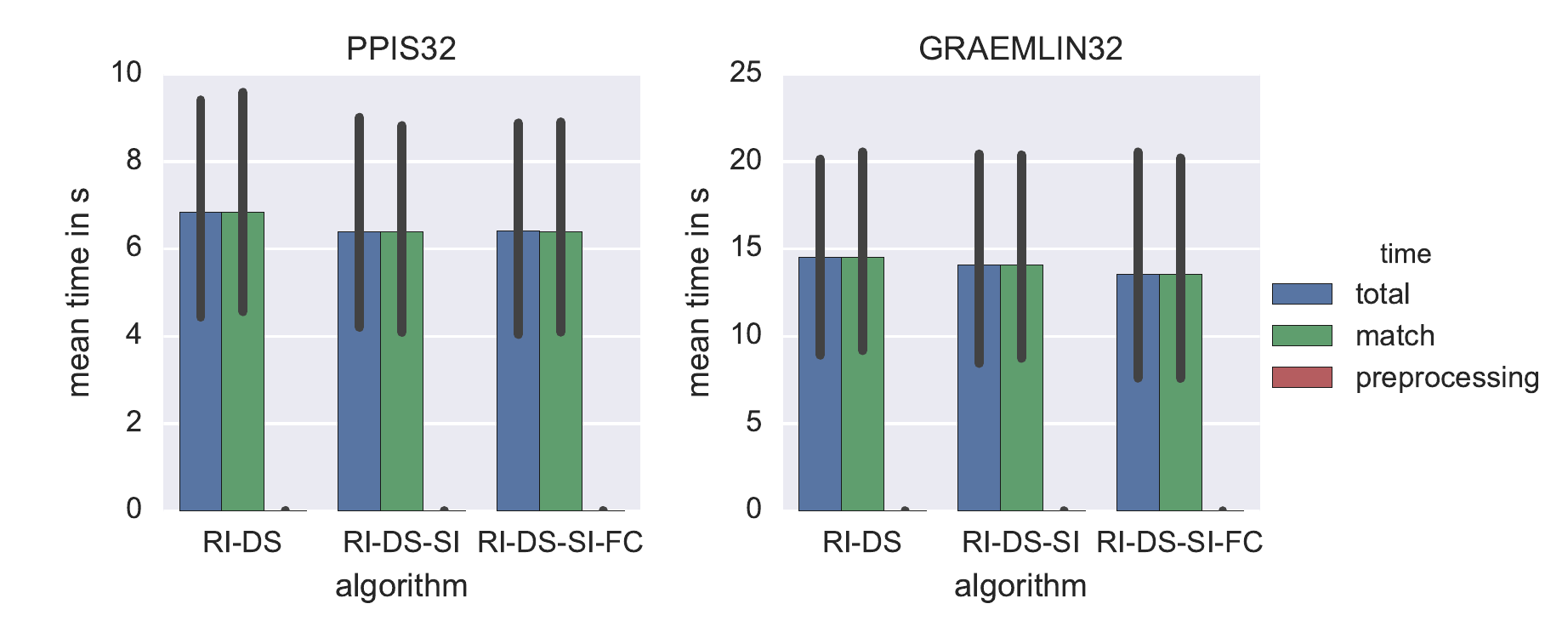}
\caption{Times on a random sample of long running instances of GRAEMLIN32
    and PPIS32 for our implementation of RI-DS in single threaded mode
    comparing different techniques.
    \imagenote{Stacked bar plots? Want plots side-by-side in single column. Might make it hard to see preprocessing.}}
\label{fig:rids_improved_slow_times}
\end{figure}
\fi


However, in \fig~\ref{fig:rids_improved_slow_search_space} we can see 
that on long running instances of PPIS32, RI-DS-SI clearly reduces the 
search space size, while RI-DS-SI-FC does not seem to affect search space
size at all. Yet, on long running instances of GRAEMLIN32, RI-DS-SI-FC shows
a clear improvement over RI-DS-SI\@. 
From \fig~\ref{fig:rids_improved_slow_times} we can see that for PPIS32, 
RI-DS-SI and RI-DS-SI-FC have similar total time and match time and both show improvements
in total time over RI-DS\@. On GRAEMLIN32, RI-DS-SI-FC beats
RI-DS-SI and RI-DS in total time, albeit not as strongly as in search space. 
Mean search space on the GRAEMLIN32 sample 
is nearly halved from RI-DS to RI-DS-SI-FC, while the mean match
time changes from 14.54 ($\sigma = 13.3$) to 13.53 ($\sigma = 13.86$).
This behavior can also be seen in \fig~\ref{fig:rids_improved_slow_search_space},
where the number of explored states per second drops between 
RI-DS to RI-DS-SI\@.  
One possible explanation for these results may be that, while we manage to reduce 
more search space, we do not change the memory access pattern
required for solving an instance.  During search, we must 
iterate over relatively short adjacency lists, implemented as arrays.
Skipping a few entries will not lead to large changes in running time, since the running time is dominated
by loading the array into memory.
Since RI-DS-SI-FC is at least as fast as RI-DS-SI, and further has a smaller
search space for the GRAEMLIN32 data collection (shown in \fig~\ref{fig:rids_improved_slow_search_space}(a)), we compare RI-DS-SI-FC with RI-DS in our remaining experiments.

\subsubsection{Parallel RI-DS-SI-FC}\label{ex:performance_parallel_rids}

\begin{figure}[]
\centering\includegraphics[width=0.75\textwidth]{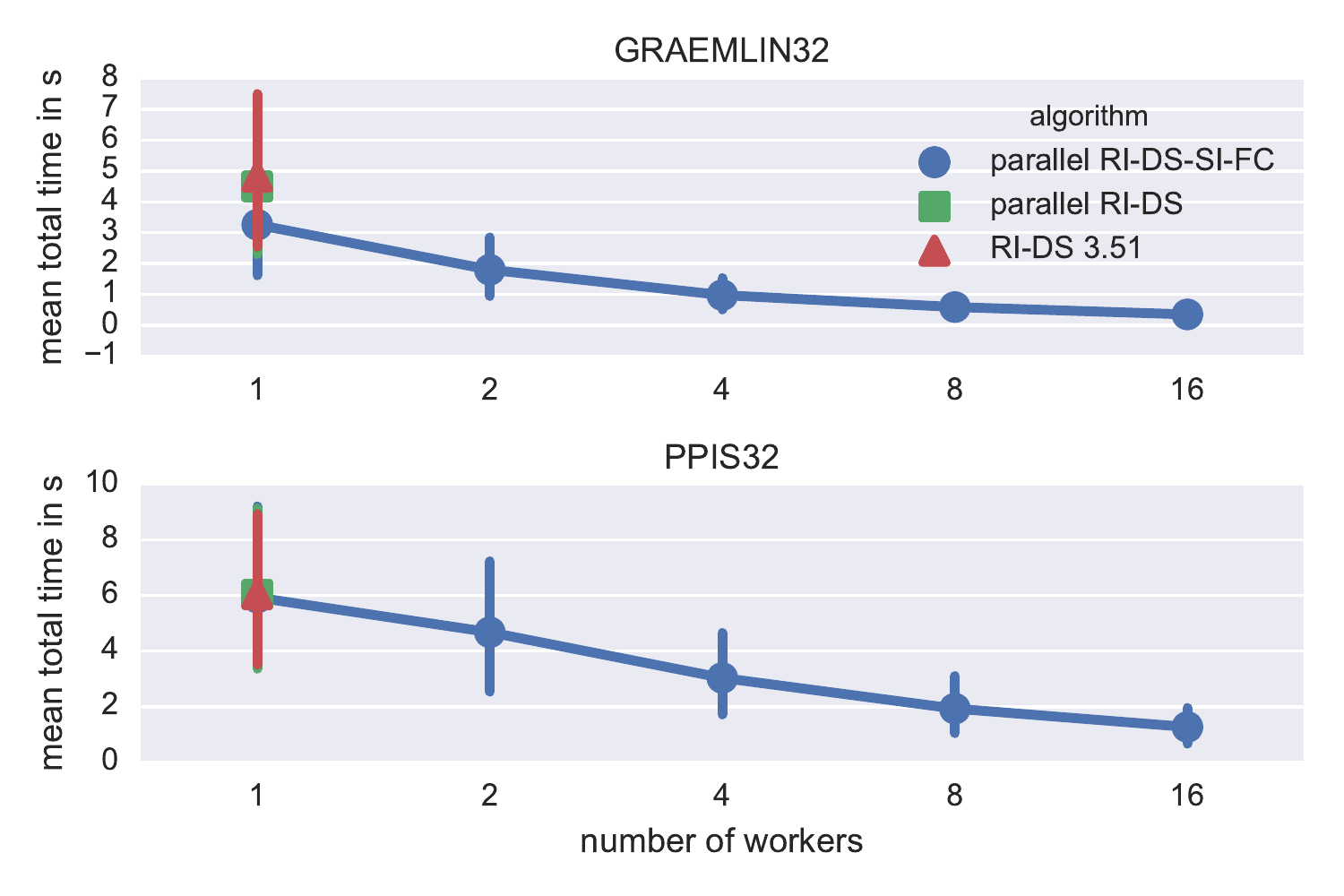}
\caption{Total time on GRAEMLIN32 and PPIS32 for variants of RI-DS.}
\label{fig:rids_time}
\end{figure}

\begin{figure}[]
\centering\includegraphics[width=0.75\textwidth]{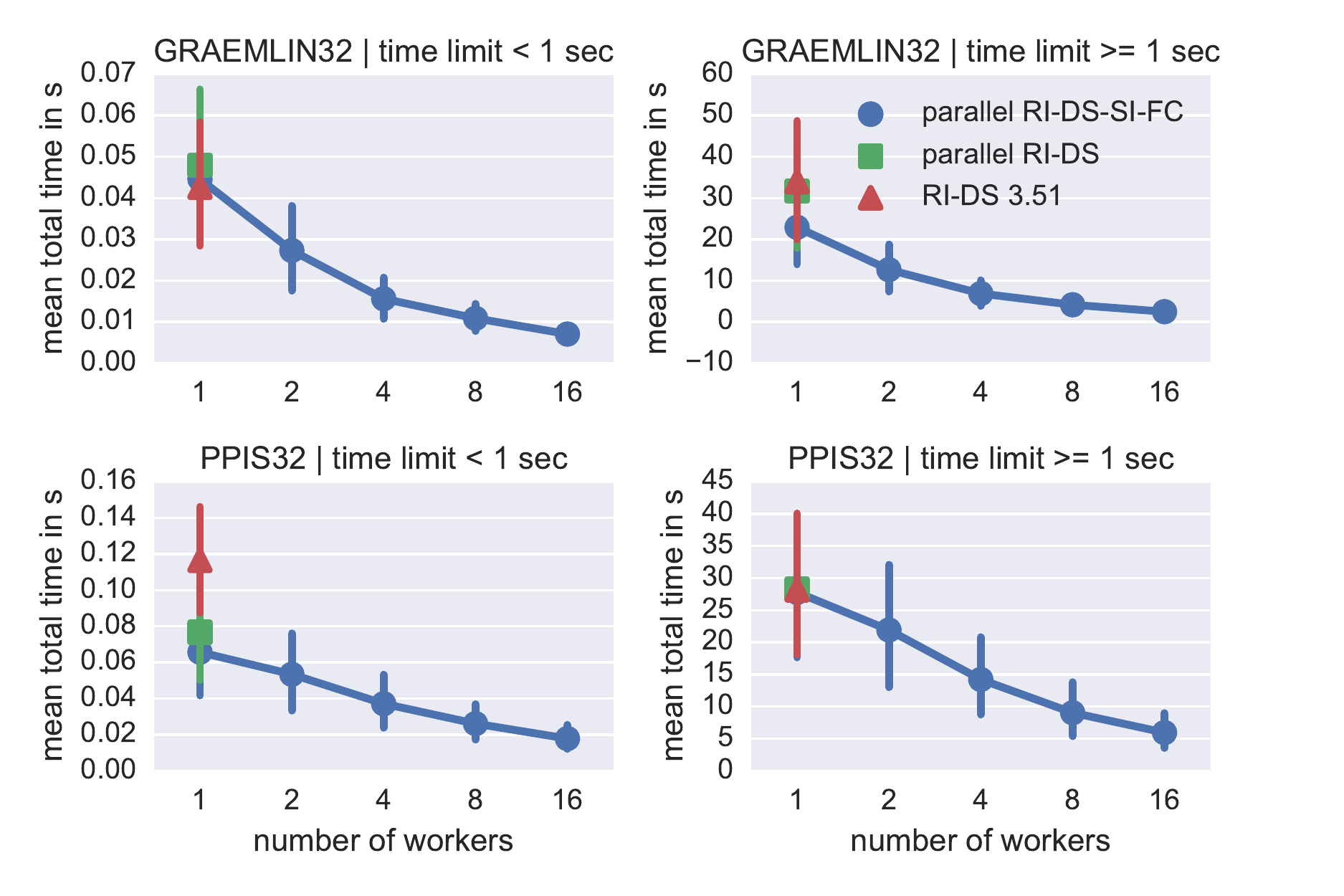}
\caption{Total time on GRAEMLIN32 and PPIS32 for variants of RI-DS split
between instances with a total time of less than one second and longer running
instances. \imagenote{Make sure to remove "collection =" from all plots, and 
give unique marks to each lineplot in a given plot.}}
\label{fig:rids_time_short_long}
\end{figure}

\ifSuggestedImage
\begin{figure}[t]
\centering
        \centering

        \includegraphics[width=\textwidth]{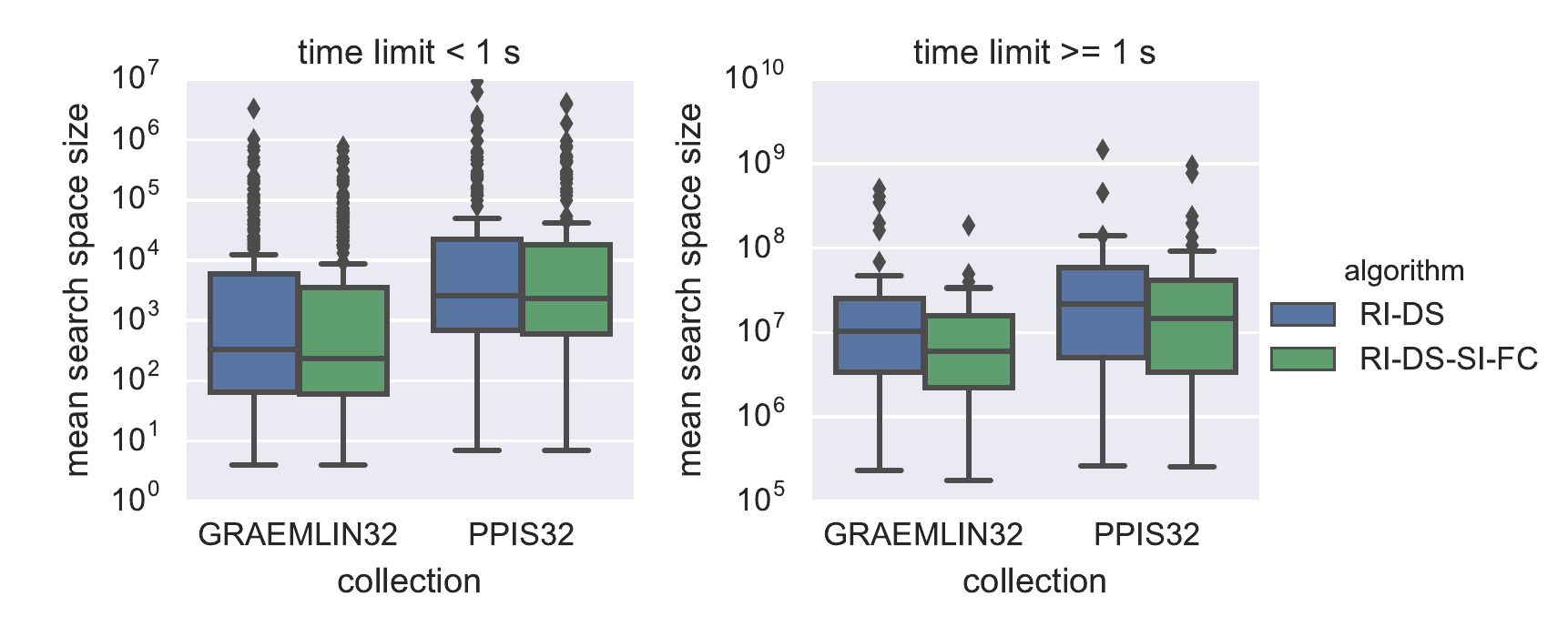}%
\caption{Search space on GRAEMLIN32 and PPIS32 for RI-DS and RI-DS-SI-FC split
between short running instances (less than one second) and longer running
instances.}
    \label{fig:rids_search_space}
\end{figure}
\else
\begin{figure*}[]
\centering\includegraphics[width=\linewidth]{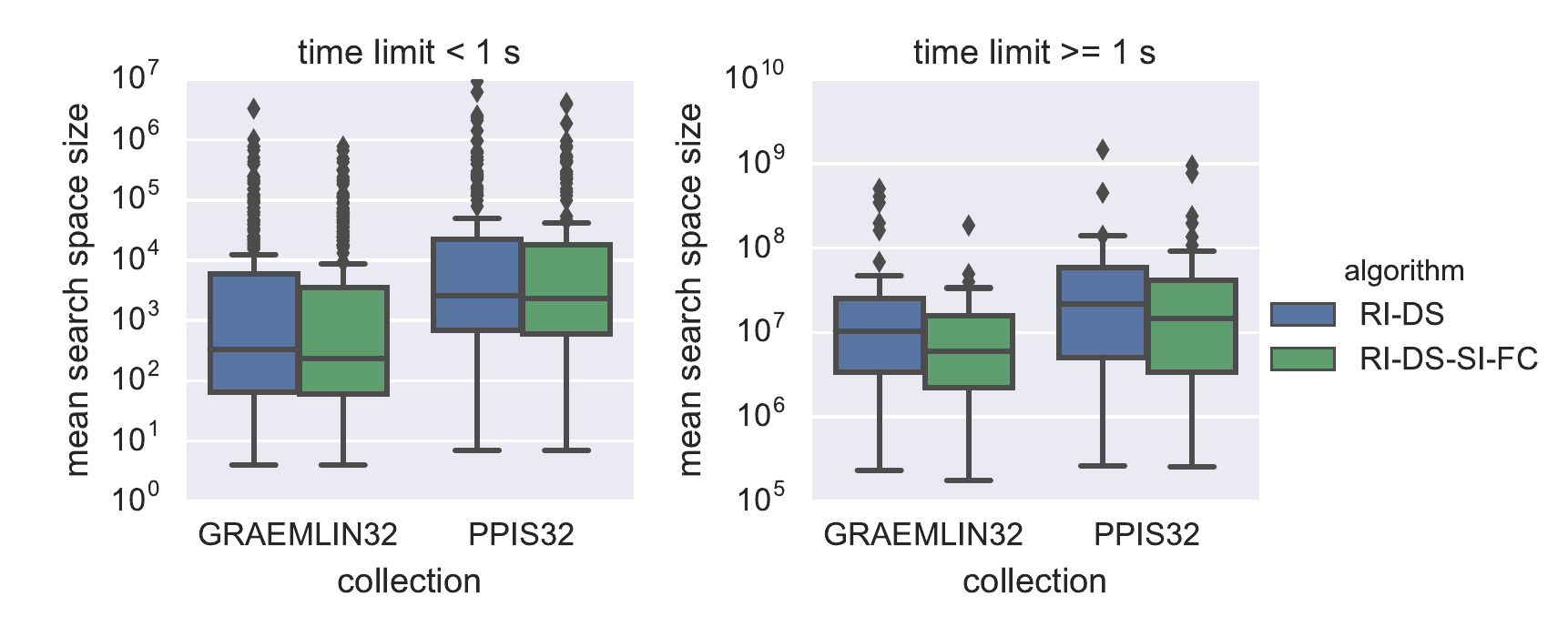}
\caption{Search space on GRAEMLIN32 and PPIS32 for RI-DS and RI-DS-SI-FC split
between instances with a total time of less than one second and longer running
instances. \imagenote{Compress into one column, use box plots. "limit = time $>=$" $\rightarrow$ time limit $\geq$ 1 sec}}
\label{fig:rids_search_space}
\end{figure*}
\fi

%
%
%

We compare parallel RI-DS-SI-FC (our improved version of RI-DS)
against our own implementation of RI-DS and the original RI-DS implementation,
RI-DS 3.51.
We measure the performance on PPIS32 and GRAEMLIN32 with 1, 2, 4, 8 and 16
workers. 
In \fig~\ref{fig:rids_time} and
\fig~\ref{fig:rids_time_short_long} we show the total time, including the time
for domain assignment and preprocessing. 
To explain the source of limited scalability, we provide not only the speedups for all instances in \fig~\ref{fig:rids_time};
in \fig~\ref{fig:rids_time_short_long} we also consider short running
instances and long running instances (with less and more than one second total
time, respectively) separately.

In \fig~\ref{fig:rids_time} and \fig~\ref{fig:rids_time_short_long}
we see that our parallelization of RI-DS (parallel RI-DS) with one
worker is slightly faster than RI-DS 3.51. 
For short running instances, RI-DS 3.5.1 is faster on GRAEMLIN32 but
slower on PPIS32. However, on long running instances, the trend is reversed; RI-DS
3.51 is slightly faster than parallel RI-DS on GRAEMLIN32 and slower on
PPIS32. We conclude that these differences are caused by implementation
differences, since the search strategy of parallel RI-DS is the same as that of
RI-DS 3.51.

As for the performance of our improved version RI-DS-SI-FC,
\fig~\ref{fig:rids_time} shows a reduction in total time of about 38\% on GRAEMLIN32 with the mean total time improving from
4.50 seconds ($\sigma = 19.25$) to 3.26 seconds ($\sigma = 14.16$). Looking
at \fig~\ref{fig:rids_time_short_long}, the improvement is especially
pronounced on long running instances. For PPIS32 the situation is
reversed. There are some improvements, but in contrast to GRAEMLIN32 on PPIS32
the numbers are improved for the short running instances, not the long running
ones. This results in a (small) improvement of about 2\% for the overall data collection with RI-DS
having a mean total time of 6.04 seconds ($\sigma = 20.45$) and RI-DS-SI-FC
having a mean total time of 5.93 seconds ($\sigma = 20.44$). 
The reduction in search space size seen in \fig~\ref{fig:rids_search_space} 
reflects this as well. The mean search space size for long running PPIS32
instances is small, while for the long running GRAEMLIN32 instances the
search space size is reduced to less than a third. Notably, for both
data collections, RI-DS-SI-FC reduces variability in search space size.
Overall these results reflect the trends in
\fig~\ref{fig:rids_improved_slow_times} where speedups for
GRAEMLIN32 on long running instances are much stronger than those for PPIS32.

\ifAppendix
\ifFull
As an aside, in Appendix~\ref{sec:appendix:rids_improved} we also give time and
search space size for RI-DS-SI-FC on PDBSv1 but as RI-DS is weaker than RI on
sparse graphs~\cite{Bonnici2013} we do not include it here. Still it should be
noted that RI-DS-SI-FC leads to noticeable improvements in search space size
and total time (about 42\%). RI-DS-SI-FC even beats RI on the long
running instances of PDBSv1. On short running instances however RI remains
much stronger.
\fi 
\fi 

\ifFull
\begin{table*}[]
   \centering
   \caption{Speedup on GRAEMLIN32 for parallel RI-DS-SI-FC with our
   implementation as baseline.}\label{tab:rids_speedup_graemlin}
   \begin{tabular}{c r c l r c l r c l}
      \toprule
      \# threads & \multicolumn{3}{c }{all instances} & \multicolumn{3}{c}{total~time~$<~1$~s} & \multicolumn{3}{c }{total~time~$\ge~1$~s}\\
      \cmidrule(lr){2-4}
      \cmidrule(lr){5-7}
      \cmidrule(lr){8-10}
      & speedup & $\min$ & $\max$ & speedup & $\min$ & $\max$ & speedup & $\min$ & $\max$ \\
      \midrule
      2  & 1.81 & 0.47 & 5.05 & 1.67 & 0.47 & 5.05 & 1.81 & 1.01 & 2.89 \\
      \midrule
      4  & 3.33 & 0.62 & 6.11 & 2.97 & 0.62 & 6.11 & 3.33 & 2.03 & 4.93 \\
      \midrule
      8  & 5.50 & 0.04 & 9.87 & 4.41 & 0.04 & 9.87 & 5.51 & 3.24 & 7.00 \\
      \midrule
      16  & 9.20 & 0.07 & 13.40 & 7.21 & 0.07 & 13.14 & 9.23 & 4.69 & 13.40 \\
      \bottomrule
   \end{tabular}
\end{table*}

\begin{table*}[]
   \centering
   \caption{Speedup on PPIS32 for parallel RI-DS-SI-FC with our implementation
   as baseline.}\label{tab:rids_speedup_ppi}
   \begin{tabular}{c r c l r c l r c l}
      \toprule
      \# threads & \multicolumn{3}{c }{all instances} & \multicolumn{3}{c}{total~time~$<~1$~s} & \multicolumn{3}{c }{total~time~$\ge~1$~s}\\
      \cmidrule(lr){2-4}
      \cmidrule(lr){5-7}
      \cmidrule(lr){8-10}
      & speedup & $\min$ & $\max$ & speedup & $\min$ & $\max$ & speedup & $\min$ & $\max$ \\
      \midrule
      2  & 1.23 & 0.44 & 2.05 & 1.21 & 0.44 & 1.97 & 1.23 & 0.59 & 2.05 \\
      \midrule
      4  & 1.97 & 0.56 & 3.82 & 1.75 & 0.56 & 3.59 & 1.97 & 1.00 & 3.82 \\
      \midrule
      8  & 3.15 & 0.07 & 6.66 & 2.64 & 0.07 & 5.82 & 3.16 & 1.69 & 6.66 \\
      \midrule
      16  & 4.83 & 0.07 & 12.98 & 4.03 & 0.07 & 9.06 & 4.83 & 2.77 & 12.98 \\
      \bottomrule
   \end{tabular}
\end{table*}
\fi

As can be seen in Table~\ref{tab:rids_speedup_ourbaseline}, speedup is greater on GRAEMLIN32 than PPIS32.
The average speedup with 16 threads on long running instances for GRAEMLIN32 is
9.49 while it is 5.21 for PPIS32. Still, both speedups are higher than the
mean speedup for parallel RI on PDBSv1 in Table~\ref{tab:ri_speedup_ourbaseline}.
The differences between data collections suggest that inherent
parallelism of the instance may be crucial for achieving
high speedup.  %
Also, the maximum speedup of 13.40 on GRAEMLIN32 shows
that, for some inputs, our implementation is capable of speedups close to the optimum.
Interestingly, while the speedup on the denser PPIS32 and GRAEMLIN32
is higher than the speedup on PDBSv1, the number of unsolved instances 
\ifFull in \fig~\ref{fig:rids_timed_out}\fi does not improve as much.
The number of unsolved instances in PPIS32 drops from 212 with RI-DS 3.51
to 197 with RI-DS-SI-FC and 16 threads and from 171 to 158 on GRAEMLIN32
respectively---suggesting that GRAEMLIN32 and PPIS32 contain many difficult
instances.

\ifFull
The overall speedup compared to RI-DS 3.51 can be seen in
Table~\ref{tab:speedup_original}. 
We
achieve the best speedup for short and long running instances of GRAEMLIN32 and
PPIS32 with 16 threads. For GRAEMLIN32 we achieve speedups of 6.67 on 
short running instances and 13.67 on long running instances.
With PPIS32 we have stronger speedup on short running instances.
We achieve a speedups of 5.12 on short
running instances and 4.37 on long running instances of PPIS32.
\else
Finally, we briefly summarize how our parallel implementation compares to RI-DS 3.51.
We achieve the best speedups for instances of GRAEMLIN32 and
PPIS32 with 16 threads. For GRAEMLIN32, we achieve speedups of 6.67 on 
short running instances and 13.67 on long running instances.
With PPIS32, we have stronger speedup of 5.12 on short running instances
when compared to 4.37 for long running instances.
\fi

\ifFull
With respect to memory usage, in \fig~\ref{fig:rids_memory} we see the same
behavior as with parallel RI\@.
Our implementation has a much larger base consumption, mostly due to more
extensive use of the C++ standard library and larger binary size.
The increase in memory usage with more threads is very moderate
and we notice that RI-DS-SI-FC has a slightly lower memory usage
than RI-DS\@.

\begin{figure}[]
\centering\includegraphics[width=\textwidth]{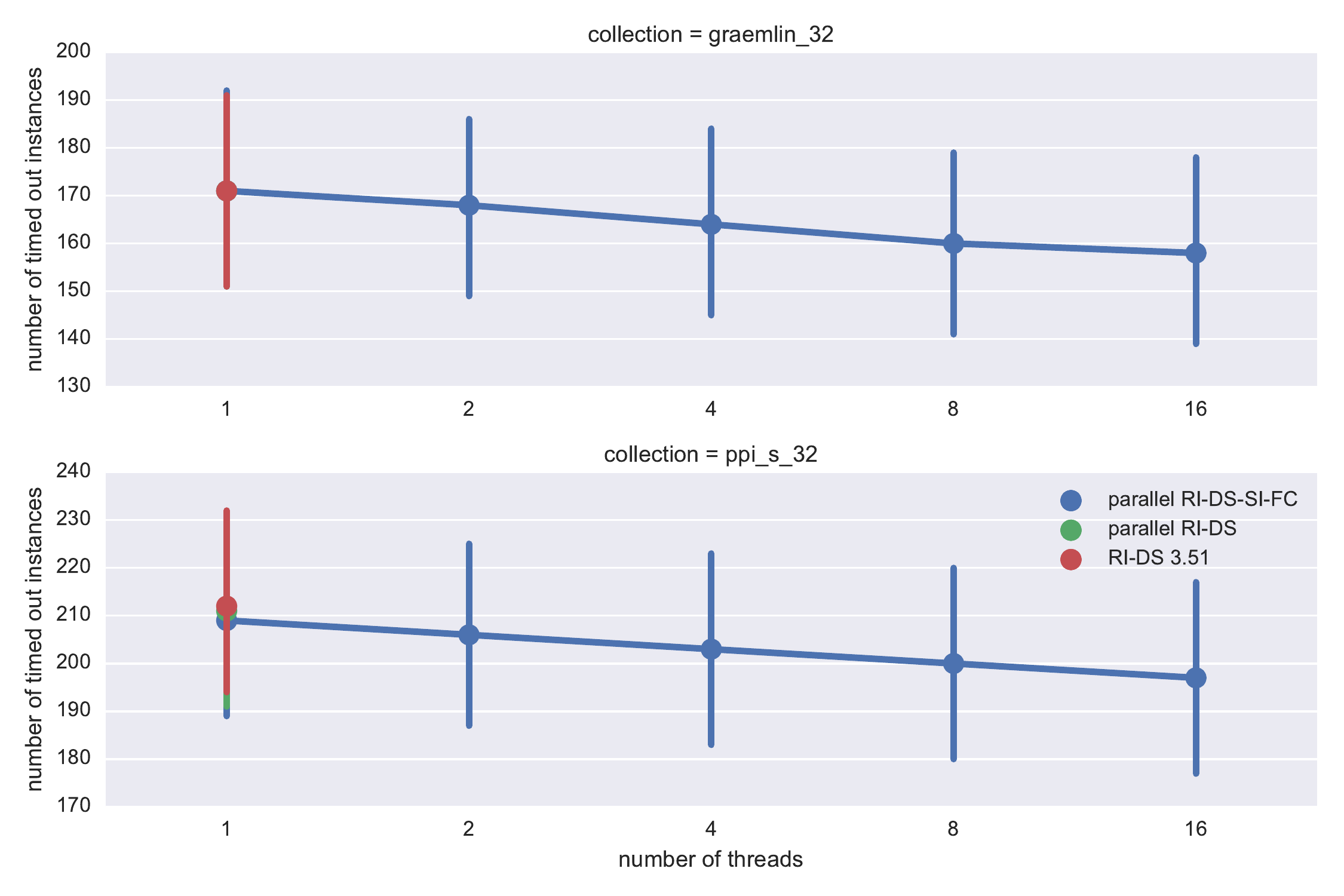}
\caption{Number of timed out instances on GRAEMLIN32 and PPIS32 for variants of RI-DS.}
\label{fig:rids_timed_out}
\end{figure}


\begin{figure}[]
\centering\includegraphics[width=\textwidth]{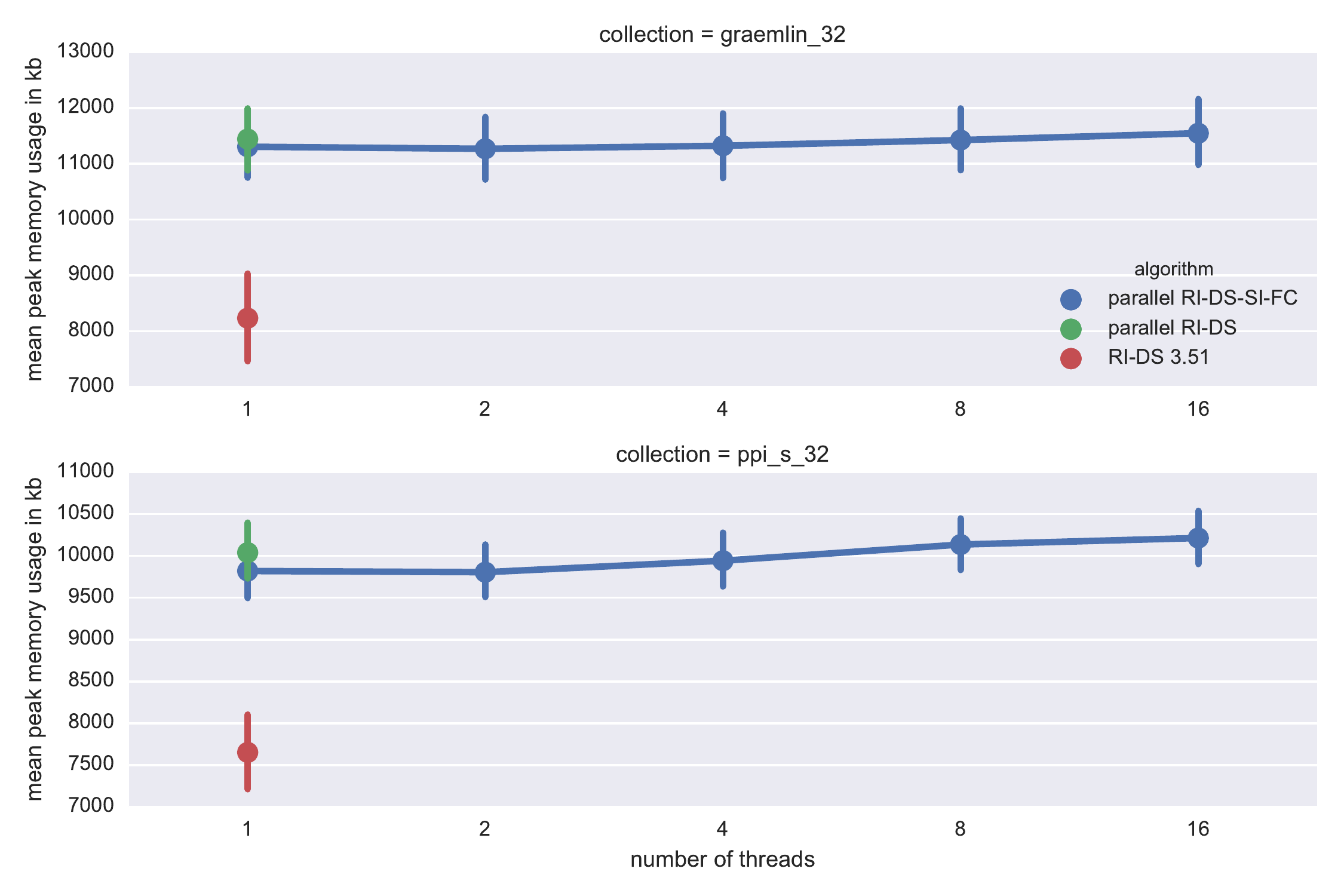}
\caption{Memory usage on GRAEMLIN32 and PPIS32 for variants of RI-DS.}
\label{fig:rids_memory}
\end{figure}
\fi

\ifOurBaseline
\ifNewNumbers
\begin{table*}[]
   \small
   \centering
   \caption{Speedup of parallel RI-DS-SI-FC over itself with one worker, for all, short, and long instances. Maximum speedups marked with a $*$ exceeded the worker count, which we attribute to cache effects for long running instances, and a lack of precision in timing for short running instances.}\label{tab:rids_speedup_ourbaseline}

   \begin{tabular}{c r r r r r r r r r }
      \toprule
      \# workers 
      & \multicolumn{3}{c }{all instances} & \multicolumn{3}{c}{short ($< 1$ sec)} & \multicolumn{3}{c }{long ($\geq 1$ sec)}\\
      \cmidrule(lr){2-4}
      \cmidrule(lr){5-7}
      \cmidrule(lr){8-10}
      & avg & gmean & $\max$ & avg & gmean & $\max$ & avg & gmean & $\max$ \\
      \midrule
      \multicolumn{7}{l}{\bf GRAEMLIN32}\\
      2  & 1.76 & 1.57 & 4.34$^*$ & 1.83 & 1.54 & 4.34$^*$ & 1.76 & 1.80 & 2.26$^*$ \\
      4  & 3.35 & 2.03 & 6.26$^*$ & 3.25 & 1.89 & 6.26$^*$ & 3.35 & 3.28 & 4.09$^*$ \\
      8  & 5.55 & 1.69 & 9.74$^*$ & 4.70 & 1.43 & 9.74$^*$ & 5.56 & 5.25 & 7.13\phantom{$^*$} \\
      16  & 9.45 & 1.38 & 13.40\phantom{$^*$} & 6.92 & 1.05 & 12.97\phantom{$^*$} & 9.49 & 8.98 & 13.40\phantom{$^*$} \\
      \midrule
      \multicolumn{7}{l}{\bf PPIS32}\\
      2  & 1.57 & 1.63 & 3.56$^*$ & 1.56 & 1.68 & 3.56$^*$ & 1.57 & 1.45 & 2.01$^*$ \\
      4  & 2.07 & 2.09 & 5.66$^*$ & 2.25 & 2.04 & 5.66$^*$ & 2.07 & 2.25 & 3.83\phantom{$^*$} \\
      8  & 3.28 & 1.85 & 7.78\phantom{$^*$} & 3.18 & 1.55 & 7.78\phantom{$^*$} & 3.28 & 3.50 & 6.74\phantom{$^*$} \\
      16  & 5.20 & 2.10 & 13.04\phantom{$^*$} & 4.80 & 1.60 & 11.27\phantom{$^*$} & 5.21 & 5.56 & 13.04\phantom{$^*$} \\
      \bottomrule
   \end{tabular}
\end{table*}

\else
\ifIncludeMin
\begin{table*}[]
   \centering
   \caption{Speedup on GRAEMLIN32 for parallel RI-DS-SI-FC with our
   implementation as baseline.}\label{tab:rids_speedup_graemlin}
   \begin{tabular}{c r c l r c l r c l}
      \toprule
      \# threads & \multicolumn{3}{c }{all instances} & \multicolumn{3}{c}{total~time~$<~1$~s} & \multicolumn{3}{c }{total~time~$\ge~1$~s}\\
      \cmidrule(lr){2-4}
      \cmidrule(lr){5-7}
      \cmidrule(lr){8-10}
      & speedup & $\min$ & $\max$ & speedup & $\min$ & $\max$ & speedup & $\min$ & $\max$ \\
      \midrule
      2  & 1.81 & 0.47 & 5.05 & 1.67 & 0.47 & 5.05 & 1.81 & 1.01 & 2.89 \\
      \midrule
      4  & 3.33 & 0.62 & 6.11 & 2.97 & 0.62 & 6.11 & 3.33 & 2.03 & 4.93 \\
      \midrule
      8  & 5.50 & 0.04 & 9.87 & 4.41 & 0.04 & 9.87 & 5.51 & 3.24 & 7.00 \\
      \midrule
      16  & 9.20 & 0.07 & 13.40 & 7.21 & 0.07 & 13.14 & 9.23 & 4.69 & 13.40 \\
      \bottomrule
   \end{tabular}
\end{table*}

\begin{table*}[]
   \centering
   \caption{Speedup on PPIS32 for parallel RI-DS-SI-FC with our implementation
   as baseline.}\label{tab:rids_speedup_ppi}
   \begin{tabular}{c r c l r c l r c l}
      \toprule
      \# threads & \multicolumn{3}{c }{all instances} & \multicolumn{3}{c}{total~time~$<~1$~s} & \multicolumn{3}{c }{total~time~$\ge~1$~s}\\
      \cmidrule(lr){2-4}
      \cmidrule(lr){5-7}
      \cmidrule(lr){8-10}
      & speedup & $\min$ & $\max$ & speedup & $\min$ & $\max$ & speedup & $\min$ & $\max$ \\
      \midrule
      2  & 1.23 & 0.44 & 2.05 & 1.21 & 0.44 & 1.97 & 1.23 & 0.59 & 2.05 \\
      \midrule
      4  & 1.97 & 0.56 & 3.82 & 1.75 & 0.56 & 3.59 & 1.97 & 1.00 & 3.82 \\
      \midrule
      8  & 3.15 & 0.07 & 6.66 & 2.64 & 0.07 & 5.82 & 3.16 & 1.69 & 6.66 \\
      \midrule
      16  & 4.83 & 0.07 & 12.98 & 4.03 & 0.07 & 9.06 & 4.83 & 2.77 & 12.98 \\
      \bottomrule
   \end{tabular}
\end{table*}
\else
\ifIncludeMax
\darren{Table~\ref{tab:rids_speedup_ourbaseline} entries are suspicious.}
\begin{table}[]
   \centering
   \caption{Speedup of parallel RI-DS-SI-FC over itself with one worker, for all, short ($<1$ sec.), and long ($\geq1$ sec.) instances.}\label{tab:rids_speedup_ourbaseline}
   \begin{tabular}{c r r r r r r}
      \toprule
      \# workers & \multicolumn{6}{c}{Speedup of parallel RI-DS-SI-FC} \\
      \cmidrule(lr){2-7}
      & \multicolumn{2}{c }{all instances} & \multicolumn{2}{c}{short} & \multicolumn{2}{c }{long}\\
      \cmidrule(lr){2-3}
      \cmidrule(lr){4-5}
      \cmidrule(lr){6-7}
      & avg & $\max$ & avg & $\max$ & avg & $\max$ \\
      \midrule
      \multicolumn{7}{l}{\bf GRAEMLIN32}\\
      2  & 1.81 & 5.05 & 1.67 & 5.05 & 1.81 & 2.89 \\
      4  & 3.33 & 6.11 & 2.97 & 6.11 & 3.33 & 4.93 \\
      8  & 5.50 & 9.87 & 4.41 & 9.87 & 5.51 & 7.00 \\
      16  & 9.20 & 13.40 & 7.21 & 13.14 & 9.23 & 13.40 \\
      \midrule
      \multicolumn{7}{l}{\bf PPIS32}\\
      2  & 1.23 & 2.05 & 1.21 & 1.97 & 1.23 & 2.05 \\
      4  & 1.97 & 3.82 & 1.75 & 3.59 & 1.97 & 3.82 \\
      8  & 3.15 & 6.66 & 2.64 & 5.82 & 3.16 & 6.66 \\
      16  & 4.83 & 12.98 & 4.03 & 9.06 & 4.83 & 12.98 \\
      \bottomrule
   \end{tabular}
\end{table}
\else
\begin{table}[]
   \centering
   \caption{Speedup of parallel RI-DS-SI-FC over itself with one worker, for all, short ($<1$ sec.), and long ($\geq1$ sec.) instances.}\label{tab:rids_speedup_ourbaseline}
   \begin{tabular}{c r r r}
      \toprule
      \# workers & \multicolumn{3}{c}{Speedup of parallel RI-DS-SI-FC} \\
      \cmidrule(lr){2-4}
      & \multicolumn{1}{c }{all} & \multicolumn{1}{c}{short ($<1$ sec.)} & \multicolumn{1}{c }{long ($\geq1$ sec.)}\\
      \cmidrule(lr){2-2}
      \cmidrule(lr){3-3}
      \cmidrule(lr){4-4}
      & avg & avg & avg \\
      \midrule
      \multicolumn{4}{l}{\bf GRAEMLIN32}\\
      2  & 1.81  & 1.67  & 1.81 \\
      4  & 3.33  & 2.97  & 3.33 \\
      8  & 5.50  & 4.41  & 5.51 \\
      16  & 9.20 & 7.21  & 9.23 \\
      \midrule
      \multicolumn{4}{l}{\bf PPIS32}\\
      2  & 1.23 & 1.21 & 1.23 \\
      4  & 1.97 & 1.75 & 1.97 \\
      8  & 3.15 & 2.64 & 3.16 \\
      16  & 4.83& 4.03 & 4.83 \\
      \bottomrule
   \end{tabular}
\end{table}
\if 
\fi 
\fi 
\fi 
\fi 

\ifFull
\begin{table}[]
\small
   \centering
   \caption{Speedup of RI-DS-SI-FC over RI-DS 3.51.}\label{tab:speedup_original}
   \begin{tabular}{r r r r}
      \toprule
      \# workers &
      \multicolumn{3}{c}{Speedup over RI-DS} \\
      \cmidrule(lr){2-4}
      & all instances & short ($<1$ sec.) & long ($\ge1$ sec.) \\

\midrule
\multicolumn{4}{l}{\bf GRAEMLIN32}\\
1 & 1.48 & 0.95 & 1.48 \\
2 & 2.67 & 1.57 & 2.68 \\
4 & 4.91 & 2.82 & 4.94 \\
8 & 8.11 & 4.17 & 8.17 \\
16 & 13.57 & 6.76 & 13.67 \\
\midrule
{\bf PPIS32} &&&\\
1 & 1.01 & 1.22 & 1.01 \\
2 & 1.28 & 1.54 & 1.28 \\
4 & 1.98 & 2.30 & 1.98 \\
8 & 3.11 & 3.35 & 3.11 \\
16 & 4.73 & 5.12 & 4.73 \\
\bottomrule
   \end{tabular}
\end{table}
\fi

\section{Conclusions}
\label{sec:conclusions}
In this paper we have presented a shared-memory parallelization for 
the state-of-the-art subgraph enumeration algorithms RI and RI-DS. 
Besides a parallelization based on work stealing, we also contribute
improved search state pruning using techniques from constraint programming.

On the long running instances of the biochemical data collections we used in our experiments,
we achieve notable (although not perfect) speedups, considering
the highly irregular data access.
We conclude that parallelization often yields a significant acceleration of state
space representation based subgraph isomorphism/enumeration algorithms, individual results
depend on the input being considered.
Future work should thus address a dynamic strategy for determining the optimal level of parallelism during
the search process.

\section*{Acknowledgment}
This work was partially supported by DFG grant ME 3619/3-1
within Priority Programme 1736.

\bibliographystyle{abbrv}
\bibliography{library}	
\end{document}